\documentclass[twocolumn,aps,prd,
preprintnumbers,
secnumarabic,amssymb,nofootinbib]{revtex4-1}
\pdfoutput=1

\usepackage{graphics}
\usepackage[dvips]{graphicx}
\usepackage{graphicx,bm}
\usepackage{mathrsfs}
\usepackage{amssymb}
\usepackage{amsmath}
\usepackage{verbatim}
\usepackage{float}
\usepackage{slashed}
\usepackage{bbm}
\usepackage{color}
\usepackage{xcolor}

\usepackage[colorlinks=false, linkbordercolor=red, citebordercolor=green, urlbordercolor=cyan]{hyperref}%

\usepackage[english]{babel}

\newcommand{\tev}{\text{ TeV}} 
\newcommand{\gev}{\text{ GeV}} 
\newcommand{\ev}{\text{ eV}} 
\newcommand{\ccmm}{\text{ cm}} 
\newcommand{\nlsp}{\widetilde{\rho}_{1}^{\pm}}
\newcommand{\nnlsp}{\widetilde{\chi}_{1}^{0}}

\begin{document}

\title{Constraining the $\boldsymbol{R}$-symmetric chargino NLSP at the LHC}

\author{Carlos Alvarado}
\email{calvara1@nd.edu}
\author{Antonio Delgado} 
\email{adelgad2@nd.edu}
\author{Adam Martin}
\email{amarti41@nd.edu}
\affiliation{Department of Physics, University of Notre Dame, 225 Nieuwland Hall, Notre Dame, Indiana 46556, USA}

\begin{abstract}
We present a phenomenological study of Dirac electroweakinos in a $U(1)_{R}$ extension of the  MSSM with a strictly $R$-symmetric Higgs sector (MRSSM) and gauge-mediated supersymmetry breaking. One of the distinguishing features of  the MRSSM is that the lightest chargino can be lighter than the lightest neutralino. Decays from the NLSP chargino to the gravitino LSP will produce exotic signals. We apply LHC-13 mass limits from both prompt and long-lived searches to the chargino NLSP regime of the MRSSM. Imposing the additional constraints coming from the 125 GeV Higgs and from  the electroweak sector, regions of the parameter space are found where the gravitino LSP, chargino NLSP scenario survives all current bounds. We also show that the fine-tuning of the model can reach a level slightly better than sub-percent with our choice of parameters.
\end{abstract}

\maketitle


\section{Introduction}
\label{sec:intro}

Weak scale supersymmetry persists as a compelling, well-motivated extension of the standard model (SM) that tackles its incompleteness in different fronts. On the purely aesthetic side, these include a possible connection with gravity \cite{Dimopoulos:1981yj,PhysRevD.13.3214}, while on the TeV physics front, supersymmetry stabilizes the Higgs vacuum expectation value (vev) from quantum corrections, comes with a built-in radiative breakdown of the electroweak (EW) symmetry, and contains a candidate for Dark Matter.

A supersymmetric extension of the SM needs to be natural, i.e. no large mass hierarchies, when one insists on employing supersymmetry as a stabilizing symmetry for the EW scale against radiative corrections \cite{Papucci:2011wy}. This is achieved through a low-scale spectrum of squarks, gluinos and light electroweakinos, which are respectively detected as jets and missing transverse energy ($\slashed{E}_{T}$) at colliders. Following this line of thought, multiple Run-I and Run-II LHC searches have looked for superpartners in the context of the Minimal Supersymmetric Standard Model (MSSM), but to date nothing has been found. The lack of signals beyond SM backgrounds translates into mass limits. Under the MSSM interpretation, the current limits are  $m_{\widetilde{q}}\geq 1.5\tev$ \cite{ATLAS-CONF-2017-022}, $m_{\widetilde{t}}\geq 1.1\tev$ (for non-compressed spectra) \cite{Sirunyan:2017xse,Aaboud:2017ayj} and $m_{\widetilde{g}}\geq1.8-1.96\tev$ from multijet plus $\slashed{E}_{T}$\cite{Sirunyan:2017cwe} or decays to third-generation squarks plus neutralinos\cite{ATLAS-CONF-2017-021}. These increasing bounds are a direct challenge to the premise of a theory without large, unnatural cancellations, as the stop mass and gluino mass feed into corrections to the soft mass of the Higgs and must be cancelled off to achieve electroweak symmetry breaking (EWSB). The higher the LHC stop and gluino limits, the larger the Higgs mass corrections, and the more finely tuned the theory. 
The situation is further complicated by the fact that the measured Higgs mass is difficult to achieve in the MSSM without large radiative corrections from heavy particles.

Motivated by these issues, several studies have abandoned minimality in favor of a better agreement with null experimental results on colored supersymmetric particles and/or less fine-tuning (FT)\cite{Fox:2002bu,diCortona:2016fsn,Arvanitaki:2013yja,Martin:2015eca,Kribs:2007ac,Choi:2010an, Kribs:2010md,Bertuzzo:2014bwa,Diessner:2014ksa,Goodsell:2015ura,Kribs:2008hq,Kribs:2013eua}. Of particular interest is the possibility of Dirac gauginos \cite{FAYET1978417}. Unlike Majorana masses, Dirac gaugino masses respect an $R$-symmetry present in the supersymmetric kinetic terms\footnote{Not to be confused with $R$-parity, $P_{R}=(-1)^{3(B-L)+2s}$.}. This seemingly small change has several profound consequences. At the very least, Dirac gaugino models require new matter fields in the adjoint representation of the gauge group; the fermionic components of these adjoint fields are what pair up with the familiar gauginos to form Dirac particles. While there are many different ways to incorporate Dirac gauginos into a supersymmetry model, one interesting possibility which we will focus on here is to impose the $R$-symmetry respected by the Dirac masses onto all other interactions. This variant of supersymmetry is known as the Minimal $R$-symmetric Supersymmetric Standard Model (MRSSM). With all interactions restricted by $U(1)_R$, it is not surprising that the behavior of the MRSSM is quite different from the MSSM. As one example, collider limits on colored sparticles in the MRSSM are significantly weaker than in the MSSM as several processes such as $pp \to \widetilde{q}_{L}\widetilde{q}_{L}, \widetilde{q}_{R}\widetilde{q}_{R}$ violate $R$-symmetry and are therefore forbidden.\footnote{In addition, Dirac gauginos naturally sit in the $\sim$ several TeV mass range, kinematically suppressing processes like $\widetilde{q}_{L,R}^{*}\widetilde{q}_{L,R}$ and the mixed-handedness modes $\widetilde{q}_{L}\widetilde{q}_{R}$, $\widetilde{q}_{L}^{*}\widetilde{q}_{R}^{*}$ with respect to the Majorana MSSM case~\cite{Kribs:2012gx, Kribs:2013oda}.} Dirac gaugino models have consequences beyond changing the character of the gaugino mass. In particular, the dependence on the supersymmetry breaking messenger scale in the gaugino one-loop correction to squark masses is removed. That property, dubbed \textit{supersoftness}, relaxes the fine-tuning of EWSB \cite{Fox:2002bu,diCortona:2016fsn,Arvanitaki:2013yja,Martin:2015eca}. 

There are multiple phenomenological and model-building studies of Dirac gaugino models in the literature. For example, $R$-symmetric Higgs sectors and the amelioration of the supersymmetric flavor problem were studied in Ref.~\cite{Kribs:2007ac}, Refs.~\cite{Choi:2010an,Chakraborty:2015gyj} examined the production of the $R$-symmetric scalars,  a viable embedding in gravity-mediated supersymmetry breaking was shown in~\cite{Kribs:2010md}, and loop level analyses of the Higgs potential were performed in Refs.~\cite{Bertuzzo:2014bwa,Diessner:2014ksa,Braathen:2016mmb}.

A peculiar feature of the MRSSM electroweakino sector is that the lightest chargino \textit{can} be lighter than the lightest neutralino. This was first demonstrated in Ref.\cite{Kribs:2008hq}, and shown to persist in a wide region of parameter space. The same reference also brought attention to the peculiar collider signals that result when a specific realization of supersymmetry breaking, gauge mediation (GMSB), is adopted \cite{Giudice:1998bp}. In GMSB, the gravitino\footnote{$\widetilde{G}$ is the spin-3/2 partner of the spin-2 graviton.}  ($\widetilde{G}$) is the lightest supersymmetric particle (LSP), and the decay of the lightest chargino (the next-to-LSP or NLSP for short) to $\widetilde{G}$ leads to a variety of distinctive exotic signals, spectrum and decay patterns that facilitate identification.

The focus of the current paper is this chargino NLSP regime of the MRSSM with gauge mediated supersymmetry breaking. Complementarily to standard jets$+\slashed{E_{T}}$ searches, we are motivated by the fact that interpretations of LHC results in terms of $R$-symmetric chargino NLSP are relatively unexplored and that up-to-date analyses of the corresponding exotic signals (\textit{displaced dijets},~\textit{disappearing} tracks and \textit{kinks}) are available \cite{Liu:2015bma,Aaboud:2017iio,Jung:2015boa}. Our goal is to apply the latest $\sqrt{s}=13\tev$ results from the LHC to bound the masses of the chargino NLSP and gravitino LSP, while reproducing  $m_{h}=125\gev$ (in its specific MRSSM realization). The topologies of our interest are those corresponding to Drell-Yan production of the light electroweakinos, such as chargino NLSP pair production or production of a chargino NLSP plus a neutralino (the next-to-next-lightest supersymmetric particle, or NNLSP). The specific constraining searches \textemdash prompt or long-lived\textemdash will depend on the scale of the gravitino mass. We will also `complete' the model by later specifying the rest of the spectrum not directly involved with the chargino NLSP signals or the Higgs mass.  Having the full spectrum then allows us to calculate the fine-tuning (FT) at a benchmark point. Even though supersoftness has demonstrated improving the FT with respect to the MSSM, the inclusion of non-supersoft operators (as will be done here) suggests that the same level of FT of purely supersoft setups may not be strictly maintained. 

The layout of the present work is as follows: the next section summarizes the MRSSM, highlighting the role of the $R$-symmetry and the new states. Next, Sec. \ref{sec:NLSPandLHC} shows the current collider constraints on the chargino NLSP mass for the prompt and exotic, long-lived cases. After that, Sec. \ref{sec:mHiggs} portrays the parametric form of the Higgs mass at one-loop and its degree of compatibility with the chargino NLSP regime.  Later, Sec. \ref{sec:spectrum} presents the rest of the model's mass spectrum, with emphasis on keeping the sfermions safe from observed limits. The computation of the FT in our class of models is first quoted analytically, and then numerically estimated in Sec. \ref{sec:FT}. Concluding remarks can be found in Sec. \ref{sec:conclusion}.



\section{The model}
\label{sec:model}

The MRSSM is a Dirac gaugino model where the $R$-symmetry preserved by the supersymmetric and gaugino masses is enforced on the entire theory. In this section we will briefly review the field content and interactions of the model.

As we want to impose an $R$-symmetry on all interactions, the first step is to identify a consistent set of $R$-charges that admits the terms we require and forbids as many dangerous operators as possible. As superpotential terms must have $R=+2$, the choice  $R[Q,U^{c},D^{c},L,E^{c}]=+1$ and $R[H_{u,d}]=0$ allows the usual MSSM Yukawa superpotential while guaranteeing that EWSB does not also spontaneously break $R$. By the same logic, gauge superfields carry $R[\mathcal{W}_{ \widetilde{B},\widetilde{W},\widetilde{g} }]=+1$ -- this charge is inherited by their fermionic components, the gauginos $\chi_a$, while the gauge fields have $R = 0$. From this charge assignment we can see that $R$-symmetry is incompatible with Majorana gaugino masses, as $R[\chi_{a}\chi_{a}]\neq 0$. The same reasoning forbids the traditional MSSM $\mu$ term, MSSM $A$-terms and dim-5 $\Delta B=1$ and $\Delta L=1$ operators. 

Gaugino and higgsino masses are a phenomenological necessity, and since they are not consistent with the $R$-symmetry from the MSSM field content alone, the model needs to be extended. Specifically, the theory is enlarged to include chiral superfields $A_{a}$ (with $a=\widetilde{B},\widetilde{W},\widetilde{g})$ respectively transforming as a singlet of hypercharge, and the adjoint representations of weak isospin  and color. These $A_{a}$  fields supply new fermion partners $\psi_{a}$ for the usual gauginos $\chi_{a}$ to form Dirac mass terms $M_{a}^{D}\chi_{a}\psi_{a}$. These mass terms respect $U(1)_{R}$ if we choose $R[A_{a}]=0$ for their parent superfield. With this charge, $R[\psi_{a}]=-1$, the exact opposite of the gaugino $R$-charge. To generate higgsino masses, we extend the field content by a pair of doublets $R_{u,d}$ carrying $R=+2$ and  with hypercharge $\mp 1/2$ so that they can form superpotential bilinears with $H_{u,d}$:
\begin{equation}
\mu_{u}R_{u}H_{u}+\mu_{d}H_{d}R_{d}~,  \label{eq:muTerms}
\end{equation}
where $\Phi \Phi\equiv \varepsilon_{ij}\Phi_{i}\Phi_{j}$ (sign convention $\varepsilon_{12}=+1$).  These doublets, however, \textit{do not} participate in EWSB (i.e. $\langle R_{u,d}^{0}\rangle=0$) which also keeps the $R$-symmetry unbroken. Note that the singlet and triplet scalar adjoints, who have null $R$-charge, can and generically do acquire vevs $v_{\tilde{B},\tilde{W}}\equiv \langle A_{\tilde{B},\tilde{W}}\rangle$ after EWSB.

While $R$-symmetry prevents us from writing Yukawa terms between $R_{u,d}$ and $Q,U^{c},D^{c},L,E^{c}$, trilinear interactions between $R_{u,d}$ the usual Higgses and the adjoints $A_a$ are possible and will prove to be important for generating a viable Higgs mass.
\begin{align}
W\supset \lambda^{u}_{\widetilde{B}} & A_{\widetilde{B}}R_{u}H_{u}+\lambda^{u}_{\widetilde{W}}A_{\widetilde{W}}R_{u}H_{u} \notag \\
&+ \lambda^{d}_{\widetilde{B}}A_{\widetilde{B}}H_{d}R_{d}+\lambda^{d}_{\widetilde{W}}A_{\widetilde{W}}H_{d}R_{d}~. \label{eq:RYukawas}
\end{align}

The $G_{\text{SM}}\times U(1)_{R}$ charges of the MRSSM matter and gauge content are summarized below in Table \ref{tab:Rcharges}. Straightforward superfield expansion gives the $R$-assignments for their bosonic and fermionic components.
\begin{table}[h!]
\centering
\begin{tabular}{ |c|c|} 
 \hline
 superfield & $U(1)_{R}$-charge \\ 
 \hline
 $Q,U^{c},D^{c},L,E^{c}$ & $+1$ \\ 
 $H_{u,d}$ & $0$ \\ 
  $\mathcal{W}_{\widetilde{B},\widetilde{W},\widetilde{g}}$ & $+1$ \\ 
 \hline
  $A_{\widetilde{B},\widetilde{W},\widetilde{g}}$ & $0$ \\ 
  $R_{u,d}$ & $+2$ \\
  \hline
\end{tabular}
\caption{$R$-charges for the MSSM fields (chiral multiplets and SM strength superfields) and the MRSSM extension (adjoints chiral superfields and doubly $R$-charged Higgses).}
\label{tab:Rcharges}
\end{table}

Having summarized the supersymmetric interactions of the MRSSM, we turn to the supersymmetry breaking effects.

\subsection{Supersymmetry breaking}
\label{sub:SUSYbreak}
Dirac gaugino masses $M_{a}^{D}$ are generated through a hidden sector $D$-type spurion $\mathcal{W}'\equiv \theta D'$ by
\begin{equation}
\int d^{2}\theta \dfrac{\mathcal{W}'\cdot \mathcal{W}_{a}}{\Lambda_{\text{mess}}}A_{a}~, \label{eq:CSO}
\end{equation}
dubbed the \textit{classical supersoft} operator \cite{Fox:2002bu}. Expanding in components, one finds that $M_{a}^{D}=\tfrac{1}{\sqrt{2}}\langle D'\rangle/\Lambda_{\text{mess}}$, where the scale $\Lambda_{\text{mess}}$ stands for the scale of communication of supersymmetry breaking. In a scheme where supersymmetry is broken by $\mathcal{W}'\equiv \theta D'$ alone, sfermion masses are generated radiatively through loops of Dirac gauginos and adjoints. Crucially, these loops are~\textit{finite},
\begin{equation}
( \bigl. m_{\widetilde{f}}^{2} \bigr )_{D}=\sum_{a}\dfrac{\alpha_{a}}{\pi}C_{a}(f)(M_{a}^{D})^{2} \log{\left( \dfrac{m_{\phi_{a}}^{2}}{(M_{a}^{D})^{2}} \right)}~, \label{eq:finiteLog}
\end{equation}
where $a$ runs over the SM factor groups, $C_{a}(f)$ is the quadratic Casimir under group $a$ for the superfield $f$, and $m_{\phi_a}$ is the mass of the scalar adjoint partner to $\psi_a$. The absence of any $\Lambda_{\text{mess}}$ dependence in $m_{\widetilde{f}}^{2}$ has two important implications. First, as the squark masses are insensitive to the largest scale in the problem, so are any quantities derived from them, such as the Higgs soft mass. Removing (reducing) the $\Lambda_{\text{mess}}$ dependence in $m^2_{H_u}, m^2_{H_d}$ translates to a significantly smaller traditional fine-tuning measure \cite{Fox:2002bu}. Second, unlike the MSSM, where running tends to erase any hierarchies between the sfermions and the gauginos, sfermion masses in Dirac supersymmetry are entirely a threshold effect. As such, $m_{\widetilde{f}} \ll M^D$ is completely natural. Plugging in some numbers, one could have a heavy ($5$-$10\tev$) gluino, consistent with LHC data, while keeping $\sim\text{TeV}$ squarks. The operator in Eq.~\eqref{eq:CSO} is the minimal ingredient for Dirac gaugino masses, and supersymmetry breaking based exclusively on it is known as the strictly supersoft limit. In this case, the sfermion physical masses are fully specified once the Dirac gaugino masses are chosen. 
 
While predictive and simple, the spectrum of strictly supersoft supersymmetry has its flaws. For one, the symmetries that allow (\ref{eq:CSO}) also permit the following $D$-breaking operator
\begin{equation}
\int d^{2}\theta \dfrac{\mathcal{W}'\cdot \mathcal{W}'}{\Lambda_{\text{mess}}^{2}}A_{a}A_{a}~\label{eq:lemon},
\end{equation}
known as the \textit{Lemon Twist} term. Expanding in components, Eq.~\eqref{eq:lemon} generates opposite sign masses for the real and imaginary parts of the scalar adjoints and thus has the potential to create a tachyon \cite{Fox:2002bu, Arvanitaki:2013yja}. Another issue in strictly supersoft supersymmetry is that the right handed sleptons are often dangerously light. Right handed sleptons receive a finite correction to their mass from bino sector loops, and the $(4\pi/\alpha_{a})^{1/2}$ hierarchy between gauginos and sfermion masses dictated by Eq.~\eqref{eq:finiteLog} means even a $1\, \text{ TeV}$ bino only generates $\tilde m_{E^c} \sim 50\, \gev$.

To remedy these flaws, the MRSSM includes a second source of supersymmetry breaking in the form of an $F$-term vev of a chiral superfield $X$. By adding $X$, we are giving up on strict supersoftness, however the theory can still be kept $R$-symmetric by choosing a suitable $R$-charge for $X$, namely $R[X] = +2$. Notice that $X$ differs from its usual gauge mediation counterpart in that it is not a singlet superfield, and therefore Majorana masses $\int d^{2}\theta X\mathcal{W}_{a}\mathcal{W}_{a}$ for the gauginos are still forbidden.\footnote{Further motivation to consider $F$-breaking is the unavoidable presence of couplings between gravity and the supersymmetry breaking sector, although the effects from this coupling are Planck-suppressed and are realized through a different spurion $\theta^{2}F'$ that is an $R$-singlet.} 

While Majorana gaugino masses cannot be constructed from $X$, several other $R$-symmetric operators involving $X$ are possible and are listed below:

\begin{itemize}
\item \textit{$\mu$ and $B_{\mu}$ terms.} The $\mu$-terms in Eq.~\eqref{eq:muTerms} originate from
\begin{equation}
\int d^{4}\theta \dfrac{X^{\dag}}{\Lambda_{\text{mess}}}(H_{u}R_{u}+H_{d}R_{d})~.
\end{equation}
Even though the usual $\mu$-term $\int d^{4}\theta XH_{u}H_{d}$ cannot be included in the superpotential, a soft $B_{\mu}$ term itself is $R$-preserving and can be written down,
\begin{equation}
\int d^{4}\theta \dfrac{X^{\dag}X}{\Lambda_{\text{mess}}^{2}}H_{u}H_{d}~. \label{eq:BmuTerm}
\end{equation}

\item \textit{Non-holomorphic masses.}  While $X$ carries $R$ charge, $X^{\dag}X$ is clearly a singlet and can be used to form familiar supersymmetry breaking operators such as non-holomorphic soft masses for the Higgs doublets, sfermions, adjoints and $R$-Higgses: 
\begin{equation}
\int d^{4}\theta \dfrac{X^{\dag}X}{\Lambda_{\text{mess}}^{2}}\Phi^{\dag}\Phi~, \label{eq:softNonHolo}
\end{equation}
where $\Phi=Q,U^{c},D^{c},L,E^{c},H_{u,d},R_{u,d},A_{a}$. As we are assuming all supersymmetry breaking is communicated via gauge mediation, these mass terms arise at the 2-loop level and are flavor diagonal. 

\item \textit{Adjoint $B$-terms.} The adjoints acquire holomorphic masses through $F$-breaking from
\begin{equation}
\int d^{4}\theta \dfrac{X^{\dag}X}{\Lambda_{\text{mess}}^{2}}A_{a}A_{a}~. \label{eq:Bterm}
\end{equation}

\item \textit{Adjoint $A$-terms.} Provided $R[X] \ne 0$, MSSM $A$-terms remain incompatible with $U(1)_{R}$, but Higgs-adjoint and pure-adjoint $A$-terms are permitted \footnote{First pointed out in \cite{Bertuzzo:2014bwa}.}
\begin{align}
\int d^{2}\theta \dfrac{X}{\Lambda_{\text{mess}}}\biggl( &A_{\widetilde{B}}A_{\widetilde{W}}\cdot A_{\widetilde{W}}~+~A_{\widetilde{B}}H_{u}\cdot H_{d} \notag \\
&+ ~A_{\widetilde{B}}H_{u}\cdot A_{\widetilde{W}}H_{d} \biggr)+\text{H.c.} \label{eq:adjTrilinear}
\end{align}
\end{itemize}

Provided the adjoint and $E^c$ masses from $X^{\dag}X$ are large enough, both of the issues discussed in the context of strictly supersoft supersymmetry can be avoided. Concocting a spectrum does still require some care, as  non-holomorphic masses $m_{A_{a}}^{2}$ and $B$-terms $B_{a}$ for the adjoint scalars still have the potential to generate a tachyon. To keep the masses of all components positive, we require $B_{a}$ terms small enough compared to $m_{a}^{2}$ so as to maintain positive the eigenvalues of the singlet-triplet block of the pseudoscalar mixing matrix (quoted in Appendix \ref{app:EWmixing}). 

The final ingredient in the theory is the gravitino $\tilde G$. The gravitino acquires a mass by absorbing the spin-1/2 mode that arises from the spontaneous breaking of local supersymmetry. $\widetilde{G}$ couples to the hidden sector via a \textit{R-singlet} chiral supermultiplet $X'$ that develops an $F$-breaking vev $\langle F' \rangle$. We emphasize that, while numerically $\langle F' \rangle \sim \langle F \rangle \sim \langle D \rangle$, they do represent differentsources of supersymmetry breaking. The mass of the gravitino is set by $\langle F' \rangle$,
\begin{equation}
m_{\widetilde{G}}=\langle F'\rangle/(\sqrt{3}{M}_{P}^{*}) \label{eq:mass32}
\end{equation}
where $M_{P}^{*}\approx 2.4\times 10^{18}\gev$ is the reduced Planck mass \cite{Martin:1997ns}. In gauge mediation, $\Lambda_{\text{mess}} \ll M_P$, so the gravitino is orders of magnitude lighter than the other superparters (whose masses go as $m \sim \langle F\rangle/\Lambda_{\text{mess}}$ or $m \sim \langle D\rangle/\Lambda_{\text{mess}}$) and is automatically the LSP.  As $X'$ is a singlet it will inevitably generate all possible soft masses, including those that break the $R$-symmetry. However, these terms will all be suppressed  by powers of $M_{P}^{*}$ and are therefore negligible compared to the $D$-term and $F$-term contributions mentioned already.



\section{Chargino NLSP and LHC constraints}
\label{sec:NLSPandLHC}

We now turn to study some interesting  signatures that distinguish the MRSSM from the MSSM, in particular the possibility of a chargino NLSP that then decays to the LSP, the gravitino.  The discussion below details the bounds coming from the LHC for this decay in order to get a glance at the available parameter space.

Before exploring the bounds, a few properties of the $R$-symmetric electroweakinos necessary for notation must be highlighted. Due to the $R$-symmetry, there are two sets of charginos that do not mix with each other, one with $R=Q$ (dubbed $\widetilde{\chi}_{1,2}^{\pm}$) and another one with $R=-Q$ (named $\widetilde{\rho}_{1,2}^{\pm}$). The $\widetilde{\chi}^{\pm}$ and $\widetilde{\rho}^{\pm}$ mixing matrices are $2\times2$: in each one a charged adjoint fermion and a charged higgsino that share the same $R$ and $Q$ charges are paired up with a charged wino and a charged $R$-fermion (their opposite-charge counterparts). Meanwhile, the (Dirac) neutralino $4\times4$ mixing matrix pairs the neutral gauginos and $R$-fermions with the neutral adjoint fermions and higgsinos. All three electroweakino mass matrices are listed in Appendix \ref{app:EWmixing}.

It is possible to gain some intuition on how the mass of the lightest chargino ends up being smaller than that of the lightest neutralino. To this end, we will assume a large value of $\tan{\beta}$ to ensure a large enough Higgs mass (see Sec. \ref{sec:mHiggs}) and small adjoint vevs $v_{a}$ ($a=\widetilde{B},\widetilde{W}$) to avoid problems with electroweak observables: the former assumption decouples the down-type higgsino, and the latter one allows us to disregard $\lambda v_{a}$ pieces when compared to $M^{D}$ and $\mu$. In these simplifying limits,
\begin{equation}
\mathcal{M}_{\widetilde{\chi}}\approx
\left(\begin{array}{cc}
M^{D} & 0 \\
         0 & \mu
\end{array}\right),~~
\mathcal{M}_{\widetilde{\rho}}\approx
\left(\begin{array}{cc}
M^{D} & O(gv/\sqrt{2}) \\
         O(\lambda v/\sqrt{2}) & \mu
\end{array}\right), \label{eq:simpleMchargino}.
\end{equation}
The lightest chargino always sits in the $\widetilde{\rho}$ sector, as the eigenvalues are repelled by the off-diagonal mass matrix elements. The higgsino/wino composition of the lightest chargino depends on the relative sizes of $M^{D}$ and $\mu$~.

In the large $\tan{\beta}$ limit and ignoring adjoint vevs, the neutralino mixing matrix takes the block diagonal form~\footnote{The $\sqrt{2}$ here originates in having normalized $\lambda_{\tilde{W}}$ (i.e. the doublet-triplet-doublet contractions) differently in Eq. (\ref{eq:RYukawas}) compared to Ref.\cite{Bertuzzo:2014bwa}~.}
\begin{equation}
\mathcal{M}_{\widetilde{\chi}^{0}}\approx 
\left(\begin{array}{cccc}
M^{D} & 0          & 0 & O(gv/2) \\
         0 & M^{D} & 0 & O(gv/2) \\
         0 &         0 & \mu & 0 \\
O(\lambda v/\sqrt{2}) & O(\lambda v/2) & 0 & \mu
\end{array}\right);, \label{eq:simpleMneutralino}
\end{equation}
Removing the decoupled state with mass $\mu$, the structure of the neutralino mass matrix is similar to the $\rho$ case. The only difference between the two matrices is the size of the off diagonal element $\propto g$, which feeds into how much the lightest eigenvalues is repelled below $\text{min}(M^{D}, \mu)$.  As the off-diagonal element is larger in the $\rho$ sector than in the neutralino sector, the lightest chargino will be lighter than the lightest neutralino\footnote{The fact that we can use the off-diagonal elements as a proxy for the mass hierarchy relies on both the neutralino and $\rho$ masses being Dirac and therefore diagonalized by bi-unitary transformations. In the MSSM, the neutralino and chargino mass matrices are diagonalized differently so comparing eigenvalues requires more work.}. Moving away from the large $\tan{\beta}$ limit and re-introducing the adjoint vevs, the hierarchy can remain, though it becomes more complicated as we introduce multiple $\lambda$ couplings, especially with relative sign differences. The chargino-neutralino hierarchy was explored numerically in detail in Ref.~\cite{Kribs:2008hq} and will be examined here in Sec.\ref{sec:mHiggs}.  Comparing this current work with Ref.~\cite{Kribs:2008hq}, a disclaimer is in order: we did not attempt to borrow the expressions for $\Delta m_{+0}\equiv m_{\text{NLSP}}-m_{\text{NNLSP}}$ used in \cite{Kribs:2008hq}. In \cite{Kribs:2008hq}, the simplifying assumptions of a common triplet and a common singlet $\lambda$ was imposed, with the goal of making it simpler to uncover the chargino NLSP parameter space. While we also seek $\Delta m_{+0}<0$, we are also interested the set of conditions that will increase the Higgs mass  (see Sec. \ref{sec:mHiggs}). As we will show, the requirements these two conditions place on the $\lambda$ couplings are not identical.

Having stated which are the two lightest sparticles ($\widetilde{\rho}_{1}^{\pm}$ from the discussion above, and $\widetilde{G}$ from the gauge-mediation embedding) one is able to describe the relevant decay process and the class of searches sensitive to it.  Assuming that all charged scalars from the Higgs-adjoint system are heavier than the $W$, the chargino NLSP decays to a gravitino through $\widetilde{\rho}_{1}^{\pm}\rightarrow W^{\pm}\widetilde{G}$. Searches at $\sqrt{s}=13\tev$ (and also $\sqrt{s}=8\tev$ ones) have set bounds on the $m_{\text{NLSP}}$-$m_{\text{LSP}}$ and the chargino $m_{\text{NLSP}}$-lifetime planes, respectively in the prompt and long-lived cases \cite{Aad:2014vma,Sirunyan:2017qaj,Sirunyan:2017lae,Jung:2015boa}. These are results whose range include near-massless $m_{\text{LSP}}$ values, thus enabling us to reinterpret their mass limits as bounds on our model's LSP $(\widetilde{G})$ and NLSP ($\widetilde{\rho}_{1}^{\pm}$).  The mass of the gravitino will be dialed over a relatively wide range starting on sub-eV up to tens of keV,  spanning both short- and long-lived regimes. In what follows, we look at these regimes one at the time. 

\subsection{Prompt regime}
\label{sub:prompt}

First, we consider the short lifetime (i.e. prompt) regime of the chargino NLSP ($\widetilde{\rho}_{1}^{\pm}$), roughly characterized by decay distances $d_{\text{decay}}\lesssim1\text{ cm}$ \cite{Feng:2010ij}, and assume there is a neutralino NNLSP ($\widetilde{\chi}_{1}^{0}$). This was argued in the discussion below Eqs.(\ref{eq:simpleMchargino}) and (\ref{eq:simpleMneutralino}), and will be numerically demonstrated once we reach Sec. \ref{sec:mHiggs}.  We will also assume that all sfermion and adjoint scalar masses are heavy compared to the light electroweakinos and play no role in $\widetilde{\rho}_{1}^{\pm}$ or $\widetilde{\chi}_{1}^{0}$ decays.  Within this setup, a crucial issue that determines the applicability of the aforementioned searches is the possibility that the NNLSP electroweakino directly decays to a gravitino (plus another SM final state $X$), thereby skipping $\widetilde{\rho}_{1}^{\pm}$. The decay directly to gravitino occurs when kinematics are such that the partial width $\Gamma(\text{NNLSP} \rightarrow \widetilde{G}X)$ is relatively large compared to the three-body $\Gamma(\text{NNLSP}\rightarrow \widetilde{\rho}_{1}^{\pm}\bar{f}f')$. Ref.~\cite{Kribs:2008hq} quantified this effect by defining the ratio
\begin{equation}
R_{\Gamma}\equiv \dfrac{ \Gamma(\text{NNLSP}\rightarrow \tilde{\rho}_{1}^{\pm}\bar{f}f') }{ \Gamma(\text{NNLSP} \rightarrow \widetilde{G}X) }~, \label{eq:Rgamma}
\end{equation}
which is proportional to $m_{\widetilde{G}}^{2}(|\Delta m_{+0}|/m_{\text{NLSP}})^{5}$ (the full dependence of $R_{\Gamma}$ is listed in Appendix \ref{app:gravitinoWIDTH}). If the ratio $R_{\Gamma}$ is large enough, the two-body partial width $\Gamma(\widetilde{\chi}_{1}^{0}\to Z\widetilde{G})$ is sufficiently suppressed and the NNLSP decay is dominated by $\widetilde{\chi}_{1}^{0}\to \widetilde{\rho}_{1}^{\pm}f\overline{f'}$ three-body modes. 

This $R_{\Gamma}$ ratio determines what final states are populated from neutralino production (either in pairs or with a chargino) and therefore dictates what searches are the most sensitive. In particular, one of the most powerful ways to bound electroweakinos is to look for chargino-neutralino (NSLP-NNLSP) production in the final states $3\,\ell + \slashed E_T$ or $\ell^+\ell^- + jj + \slashed E_T$. However, if $R_{\Gamma}$ is large, chargino-neutralino in our scenario will populate a different final state: $pp \to \nnlsp\,\nlsp \to \widetilde{\rho}_{1}^{\mp}\,\nlsp\,f'f \to W^+W^-\, f'f + \slashed E_T$. 

If $R_{\Gamma}$ is small,  direct $\text{BR}(\widetilde{\chi}_{1}^{0}\to Z\widetilde{G})$ cannot be neglected compared with the three-body branching fractions, and searches based on $\widetilde{\chi}_{1}^{0}\widetilde{\rho}_{1}$ production that decay into $WZ$ will apply\footnote{$R_{\Gamma}$ is proportional to  the fifth-power of the NNLSP-NLSP mass difference, therefore $R_{\Gamma}$ will automatically be small in compressed scenarios.}. The most stringent of these is the $\sqrt{s}=13\tev$ search of $\ell^{+}\ell^{-}+2j+\slashed{E}_{T}$ that sets $m_{\text{NLSP}}\gtrsim 610\gev$ at a nearly massless LSP~\cite{Sirunyan:2017qaj}.

On the other hand, when $R_{\Gamma}$ is large, the fate of the scenario depends on the mass splitting between the NNLSP and the NLSP. If the mass splitting is large, the extra fermions in the $\widetilde{\chi}_{1}^{0}\to \widetilde{\rho}_{1}^{\pm}f\overline{f'}$ decay are energetic and may be captured by $3\,\ell + \slashed E_T$ or $\ell^+\ell^- + jj + \slashed E_T$ searches despite the unusual $\widetilde{\chi}_{1}^{0}$ decay. However, if the NNLSP-NLSP mass splitting is small, the extra fermions are too soft and a different search channel is needed. We will see in Sec.~\ref{sec:mHiggs} that MRSSM setups which reproduce the Higgs mass fall into this near-degenerate category, with mass splittings $\mathcal O(10\, \text{GeV})$, therefore we will focus on the small splitting scenario here.

When the electroweakino spectrum is compressed, production of any pair electroweakinos ($\widetilde{\chi}_{1}^{0}\widetilde{\rho}_{1}^{+}, \widetilde{\chi}_{1}^{0}\overline{\widetilde{\chi}_{1}^{0}}$, etc.) is indistinguishable from chargino pair production (as all other particles produced in the cascade are soft). Further, while electric charge conservation alone allows both types of chargino in the neutralino decay, $\widetilde{\chi}_{1}^{0}\to \widetilde{\rho}_{1}^{\pm}f\overline{f'}$, $R$-symmetry does not; the neutralino decays to one and the antineutralino to the other. As a result, the chargino pair resulting from any electroweakino production in the MRSSM always have the opposite sign and thus, after the charginos decay,  all fall into the $W^+W^- + 2\widetilde{G}$ final state. Beyond the standard model production of $W^+W^- + \slashed{E_T}$ is subject to opposite-sign dilepton plus MET searches. Using this channel, the ATLAS collaboration reports a wino-like chargino mass limit of $180$ GeV at $95\%$ C. L. limit \cite{Aad:2014vma}. If we assume the signal efficiency is constant and incorporate $W$ branching fractions to leptons, we can recast this limit into a `model-independent' cross section limit of:

\begin{equation}
\sigma(pp\rightarrow W^{+}W^{-})\lesssim 600\text{ fb} \label{eq:WWxsection}
\end{equation}
from 8 TeV data. Calculating cross sections in the MRSSM via \texttt{Madgraph 5} with the NNLSP-NLSP splitting fixed to $|\Delta m_{+0}|=10 \gev$ and summing over all processes that lead to $W^+W^- + 2\widetilde{G}$, we find that the above cross section limit translates into a mass limit of $m_{\widetilde{\rho}_{1}}^{\pm}\geq 220\text{ GeV}$.

Having pinpointed the mass bounds in the limiting cases of large or small $R_{\Gamma}$, it is natural to ask how the bounds interpolate between the extremes as $R_{\Gamma}$ is varied (for fixed chargino neutralino mass splitting). Stated another way, we would like to know the smallest $R_{\Gamma}$ (see Eq.~\eqref{eq:Rgamma}) such that the bound from Ref.~\cite{Sirunyan:2017qaj} applies. The actual limit placed by Ref.~\cite{Sirunyan:2017qaj} is on the product $\sigma(pp\rightarrow \widetilde{\chi}_{1}^{0}\widetilde{\rho}_{1})\text{BR}(\widetilde{\chi}_{1}^{0}\to Z\widetilde{G})$, and by estimating the production cross section at $610\gev$ we find the bound translates to
\begin{equation}
\sigma(pp\rightarrow \widetilde{\chi}_{1}^{0}\widetilde{\rho}_{1})\text{BR}(\widetilde{\chi}_{1}^{0}\to Z\widetilde{G})\leq 0.007\text{ pb}~. \label{eq:sigmaOverR}
\end{equation}
At any given chargino mass $\lesssim 610\, \gev$, we can use the Eq.~\eqref{eq:sigmaOverR} to solve\footnote{Also obtained with \texttt{Madgraph 5} and the MRSSM modelfile.} for the minimal value of $R_{\Gamma}$.

\begin{figure}[h!]
\centering 
\includegraphics[scale=0.65]{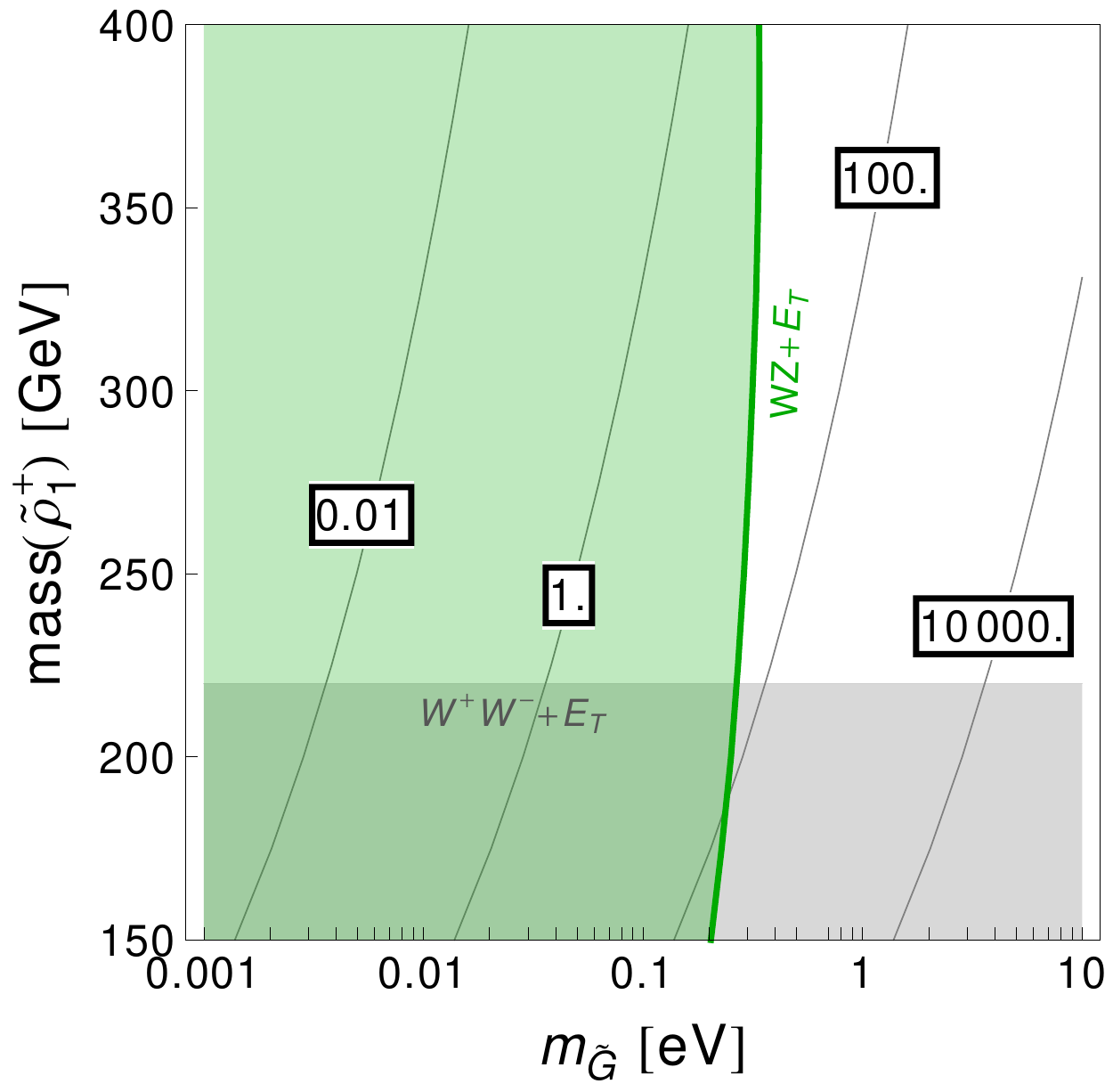}
\includegraphics[scale=0.65]{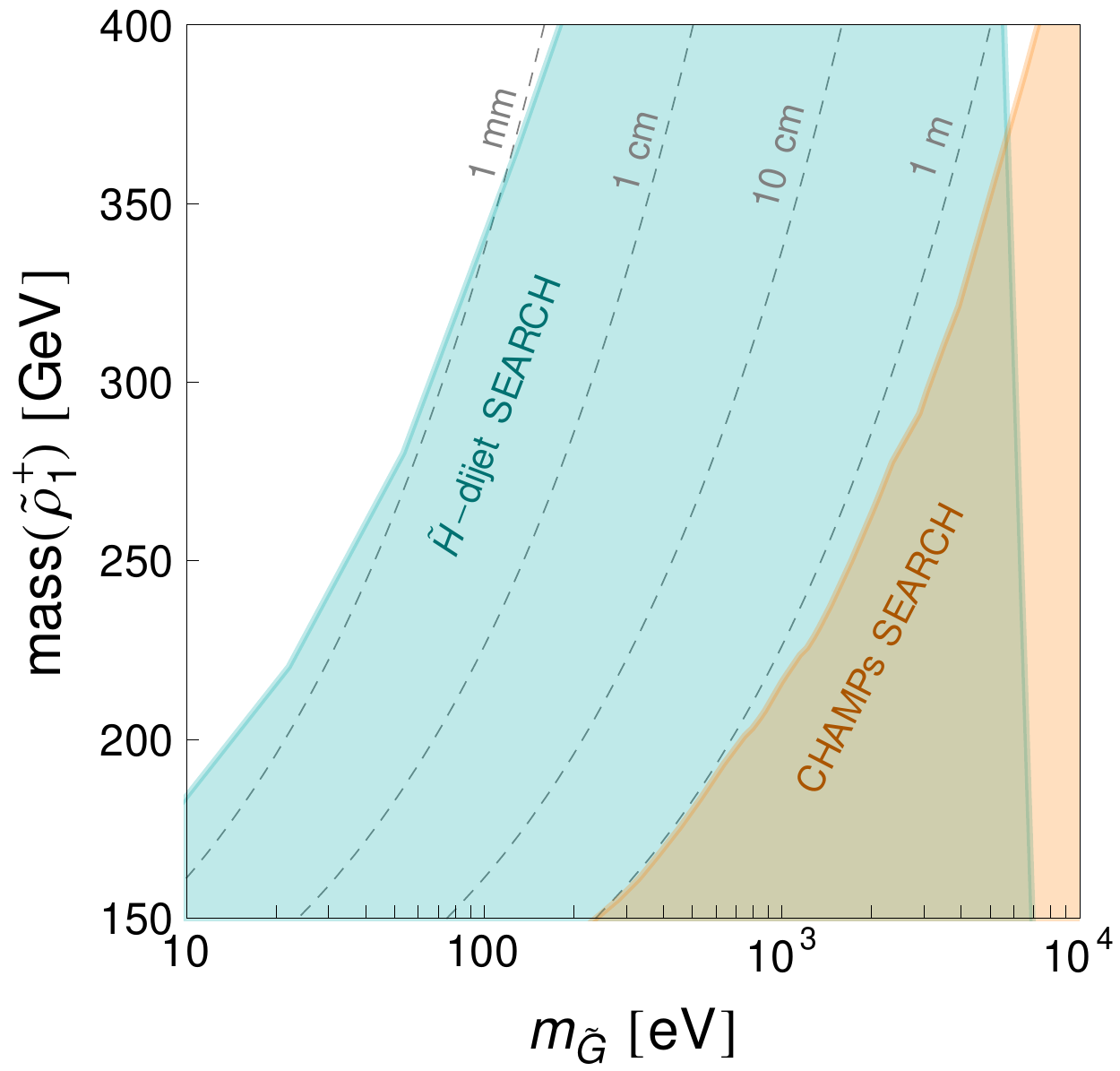}
\caption{ \textbf{Top panel:} Chargino NLSP mass constraints in the prompt (sub-eV gravitino) region with $R_{\Gamma}$ contours superimposed, for a $10\gev$ NNLSP-NLSP mass difference. For large enough $R_{\Gamma}$ only the $W^{+}W^{-}+\slashed{E}_{T}$ (gray region) bound applies. Below a minimal $R_{\Gamma}$ the $WZ+\slashed{E}_{T}$ sets the stringent constraint of Eq.(\ref{eq:sigmaOverR}) (green region). \textbf{Bottom panel:} Same mass constraints, now for macroscopic lengths (i.e. $>10\ev$ gravitino masses) from mostly-higgsino displaced dijets (blue) and long-lived CHAMPs (orange). Dashed gray contours provide reference transverse distances for decay across the detector.}
\label{fig:massGmassNLSP}
\end{figure}

To visualize how these limits impact our model, we place them on the gravitino-NLSP mass plane in the top panel of Fig.~\ref{fig:massGmassNLSP}. Contours of $R_{\Gamma}$,  indicated in Fig. ~\ref{fig:massGmassNLSP} by gray curves, have been superimposed for reference and vary with $m_{\widetilde{G}}$ according to Eq. (\ref{eq:widthChi}). For these contours, we fix $|\Delta m_{+0}|\approx 10\gev$, a mass splitting value we will show in Sec.\ref{sec:mHiggs} is characteristic of points where $m_{h}$ takes its correct value. In Fig.~\ref{fig:massGmassNLSP}, the gray shade indicates the $\ell^{+}\ell^{-}+\slashed{E}_{T}$ exclusion (from $W^{-}W^{+}$), which is independent of $R_{\Gamma}$ and thus independent of the gravitino mass, and the region in green denotes the $\ell^{+}\ell^{-}+2j+\slashed{E}_{T}$ exclusion (from $WZ$). If we had chosen smaller (larger) $|\Delta m_{+0}|$, the exclusion curve from $\ell^{+}\ell^{-}+2j+\slashed{E}_{T}$ would have the same shape but would shift to the right (left).

To summarize, the mass constraints on our model from prompt searches inform us that the gravitino cannot be too light (sub-eV), with the actual value of the bound depending on the value of the $\sim$ several-hundred GeV chargino NLSP mass. The near-verticality of the green border in Fig. \ref{fig:massGmassNLSP} implies that the size of the lower $m_{\widetilde{G}}$ bound  is roughly maintained across $m_{\text{NLSP}}$ variations of several hundred GeV.

\subsection{Long-lived regime}
\label{sub:long}

We now proceed with bounds on long-lived chargino NLSP -- charginos which decay inside the detector but far away from the primary vertex. The final state is again  $W^{\pm}\widetilde{G}$, although with macroscopic lifetimes for $\widetilde{\rho}_{1}^{\pm}$. A conventional way of classifying the possible signatures is based on our capability of detecting the chargino daughter that carries away the electric charge. The so called \textit{dissapearing tracks} (charged daughter is too soft) and \textit{kink tracks} (charged daughter is visible) belong to this classification.

An extensive recast of long-lived superpartner searches can be found in Ref. \cite{Liu:2015bma}.  One of their categories is a simplified GMSB scenario with $\sim10\ev$ quasidegenerate higgsinos, that decay though $\widetilde{H}^{0}\rightarrow Z\widetilde{G}$ . Using the exclusions set by the CMS displaced dijet analysis \cite{CMS:2014wda} ($100~\mu\text{m} \lesssim d_{\text{decay}}\lesssim 60\text{ cm}$), NLSP mass limits as stringent as $600\gev$ were set for a $O(10)\ccmm$ travel length. We use this  result as a limit on the hadronic decays of the $W^{\pm}$ pair of our own setup, an approximation that is enough given the similarity in the NLSP composition (higgsino) and the small mass difference between the $W$ and $Z$.

To translate the chargino NLSP lifetime axis of that reference into a gravitino mass axis, one makes use of the $W\widetilde{G}$ partial width in Eq. (\ref{eq:widthChi}),
\begin{equation}
\Gamma(\widetilde{\rho}_{1}^{\pm}\to W^{\pm}\widetilde{G}) =\kappa_{\widetilde{G}W}\dfrac{m_{\tilde{\rho_{1}}}^{5}}{96\pi M_{P}^{*2}m_{\widetilde{G}}^{2}}\left[ 1-\dfrac{M_{W}^{2}}{M_{\tilde{\rho_{1}}}^{2}} \right]^{4},
\end{equation}
and ends up with the constraint shown as the blue-shaded region in the bottom panel of Fig. \ref{fig:massGmassNLSP}. 
It must be pointed out that the 13 TeV analysis of Ref. \cite{Aaboud:2017iio}, that looks for long lived gluinos, could be recasted for our situation. However we do not expect that it will give any further constrain in the region shown in Fig. \ref{fig:massGmassNLSP} due to the large MET requirement ($>250$ GeV).\footnote{Heavier charginos will likely be constrained by the 13 TeV search, though these scenarios are less natural (See Sec.~\ref{sec:FT}) and therefore outside of the scope of this paper. }

For even larger lifetimes, the searches rely on identification of charged massive particles (CHAMPs) that escape the detector without  decaying ($d_{\text{decay}}>5$-$10\text{ m}$). These states propagate with high momentum, $v/c<0.9$, and high rates of ionization energy loss $dE/dx$ \cite{Khachatryan:2016sfv}. The limits \cite{Khachatryan:2015lla} on long-lived charginos masses were quoted in the mass-lifetime plane of Ref.~\cite{Jung:2015boa}, ruling out objects with $c\tau \geq1\text{ m}$. Just as done for the displaced dijet constraint, we express the CHAMPs limit in terms of $m_{\widetilde{G}}$ and depict it as the orange shaded region in the bottom panel in Fig. \ref{fig:massGmassNLSP}. 

Combining the results from both panels, we see that light chargino NLSPs are fairly constrained. For the range of chargino masses we are interested in, the bound on prompt scenarios is governed by the NLSP-NNLSP splitting and gravitino mass. For $10 \text{ GeV}$ mass splitting, we find $\widetilde{\rho}^{\pm}_1$ must be heavier than $\sim 225 \text{ GeV}$ and decay to a gravitino heavier than $0.3 \text{ eV}$. The upper limit on the gravitino mass ($\propto \widetilde{\rho}^{\pm}_1$ lifetime) is more dependent on the chargino mass, ranging between $30\, \text{eV}$ for $\widetilde{\rho}^{\pm}_1 \sim 225\, \text{GeV}$ and increasing to $100 \text{ eV}$ for $\widetilde{\rho}^{\pm}_1 \sim 350 \text{ GeV}$.

Having reviewed the regions where the chargino NLSP mass can sit according to collider data, we now shift our discussion to the Higgs mass constraint. 


\section{A $\boldsymbol{125\gev}$  Higgs}
\label{sec:mHiggs}

The presence of Higgs-adjoint mixing and the absence of NMSSM-like terms $A_{\widetilde{B}}H_{u}H_{d}$ or $H_{u}\cdot A_{\widetilde{W}}H_{d}$ in the superpotential (by the $R$-symmetry) that could help increase the Higgs quartic imply a MRSSM tree-level Higgs mass bounded by $m_{Z}$ (see Appendix \ref{app:Hmixing}). As such, we need to rely on one-loop contributions to reach $m_h = 125\gev$. Numerous parameters enter into the effective scalar potential, and a subset of them that also appear in the electroweakino masses ($M^{D}$, $\mu$, the $\lambda$'s, and $\tan \beta$) which we explored and constrained in the preceding section. Thus, the remaining step is to understand the implications for $m_{h}$ from the electroweakino constraints, and find how much flexibility is offered by the rest of the input parameters that are unrelated to the electroweakinos (i.e. the adjoint $B_{a}$, the $m_{a}^{2}$, and the stop mass). We will see that, even though large chunks of the $m_{\widetilde{G}}-m_{\tilde{\rho}_{1}^{\pm}}$ plane are disfavored by the Higgs mass condition in both chargino NLSP regimes, it is possible to accommodate $m_{h} = 125\, \gev$.

In order to comply with a $125\gev$ Higgs, we start by fixing the relevant parameters for $m_{h}$ and by reminding the reader of the required machinery to reproduce it. The enlargement of the Higgs sector field content in the MRSSM and the presence of new operators affect the way EWSB and a $125\gev$ Higgs mass are realized. Nevertheless, the vanishing vevs of the $R$-Higgses and the tiny adjoint vevs (motivated below from electroweak precision tests) dictate that EWSB is achieved much like in the MSSM \textemdash that is, radiatively \textemdash through a suitable choice of soft masses and a value for $\tan \beta$ large enough to decouple the states other than the up-type doublet.  Several references have worked the Higgs mass out at 1-loop in the strictly $R$-conserving case \cite{Bertuzzo:2014bwa,Diessner:2014ksa}; their findings for the Higgs potential are listed and commented upon below.

\begin{itemize}

\item{} \textit{$F$-terms:}  the $V_{F}$ potential is schematically given by terms
\begin{equation*}
\sum_{\Phi} \bigl| \partial_{\Phi}(\lambda AHR+\mu HR) \bigr|^{2},~~~\Phi=H,R,A~. 
\end{equation*}
from Eqs. (\ref{eq:muTerms}) and (\ref{eq:RYukawas}). After EWSB and nonzero adjoint vevs, Higgs-adjoint mixings proportional to $\lambda vv_{a}$ and to $\lambda v\mu$ are respectively induced by $|\lambda AH|^{2}$ and by the cross term in $|\lambda AH+\mu H|^{2}$ (here $v^{2}\equiv \langle H_{u}^{0}\rangle^{2}+\langle H_{d}^{0}\rangle^{2}$). Notice that insisting on $\langle R_{u,d}^{0}\rangle=0$ prevents mixing of $R_{u,d}^{0}$ with $A_{a}$ and $H_{u,d}^{0}$. Also, electroweak precision tests require the $v_{a}$ be $O(\text{GeV})$ \cite{Bertuzzo:2014bwa,Diessner:2014ksa}, implying that their effect on Higgs-adjoint mixing must be tiny compared to $\lambda v\mu$.

\item{} \textit{$D$-terms:} in addition to the MSSM Higgs $D$-term and mixed quartics with the $R_{u,d}$, the $D$-term potential contains triscalar interactions $|H_{u,d}|^{2}A_{a}$ proportional\footnote{More precisely, the triscalar couplings are proportional to the hidden sector $D$-spurion associated with $M_{a}^{D}$.} to $g_{1}M_{\widetilde{B}}^{D}$ and $g_{2}M_{\widetilde{W}}^{D}$. These originate from the supersoft operator (\ref{eq:CSO}), and they mix the Higgs with the adjoints by an amount $g_{a}vM_{a}^{D}$ after EWSB. The mixing  controlled by $g_{a}vM_{a}^{D}$ is expected to compete with the one coming from $\lambda v\mu$, especially if their signs happen to be opposite. Here $M_{a}^{D}$ will range from hundreds of GeV to about 1 TeV (depending on the gaugino scale) but they are multiplied by gauge couplings, whereas the few-hundred GeV $\mu_{u,d}$ mass parameters (responsible for higgsino mass size) multiply couplings $\lambda$ which, as shown later, will be required to be large ($\sim 1$).

\item{} \textit{Soft terms:} $V_{\text{soft}}$ includes a $B_{\mu}$ term, necessary for EWSB, in addition to two soft masses for the doublets. As opposed to their corresponding superpotential analogs, the supersymmetry-breaking triscalars\footnote{Do not mistake the adjoints $A_{a}$ for triscalar $A$-term couplings.} $A_{\widetilde{B}}H_{u}\cdot H_{d}$ and $H_{u}\cdot H_{d}A_{\widetilde{W}}$ of Eq. \ref{eq:adjTrilinear} are $R$-invariant and permitted. Nevertheless, the GMSB embedding makes these greatly suppressed, so they barely contribute to Higgs-adjoint mixing.

\end{itemize}

To determine the $m_{h}$ value at each $(M^{D},m_{\text{adj}})$ point, the parameters $\tan \beta$, $\mu$, $B_{\mu}$ and each of the $\lambda$'s must be fixed. To get started, $\tan \beta=50$ is picked from now on because a large $\tan \beta$ saturates the tree-level $(m_{h}^{2})_{\text{tree}}$ bound. Recall from the discussion on the electroweakino mass matrices (\ref{eq:simpleMchargino} and \ref{eq:simpleMneutralino}) that the composition of the chargino NLSP depends on the ordering of $\mu$ and $M^{D}$. To be consistent with the higgsino-like chargino NLSP bounds described in Sec. \ref{sub:long}, we pick these mass parameters such that $\mu<M^{D}$. Let's then adopt a reference, common $\mu_{u}=\mu_{d}\equiv \mu$ value of $\mu=250\gev$ and make sure of scanning over a $M^{D}$ ranges larger than several hundreds of GeV. A $B_{\mu}=(400 \gev)^{2}$ not too far from $\mu$ is fixed too.

\begin{figure}[h!]
\centering 
\includegraphics[scale=0.65]{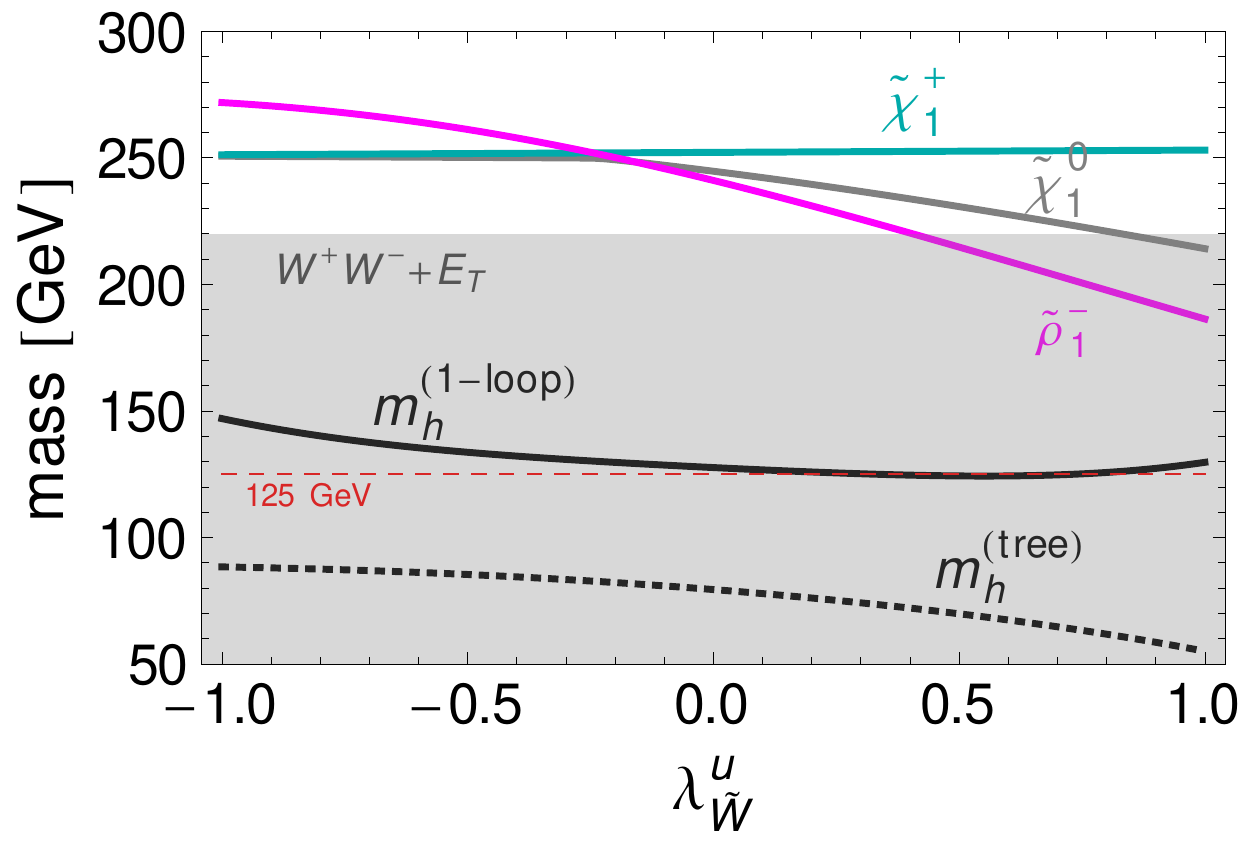}
\includegraphics[scale=0.65]{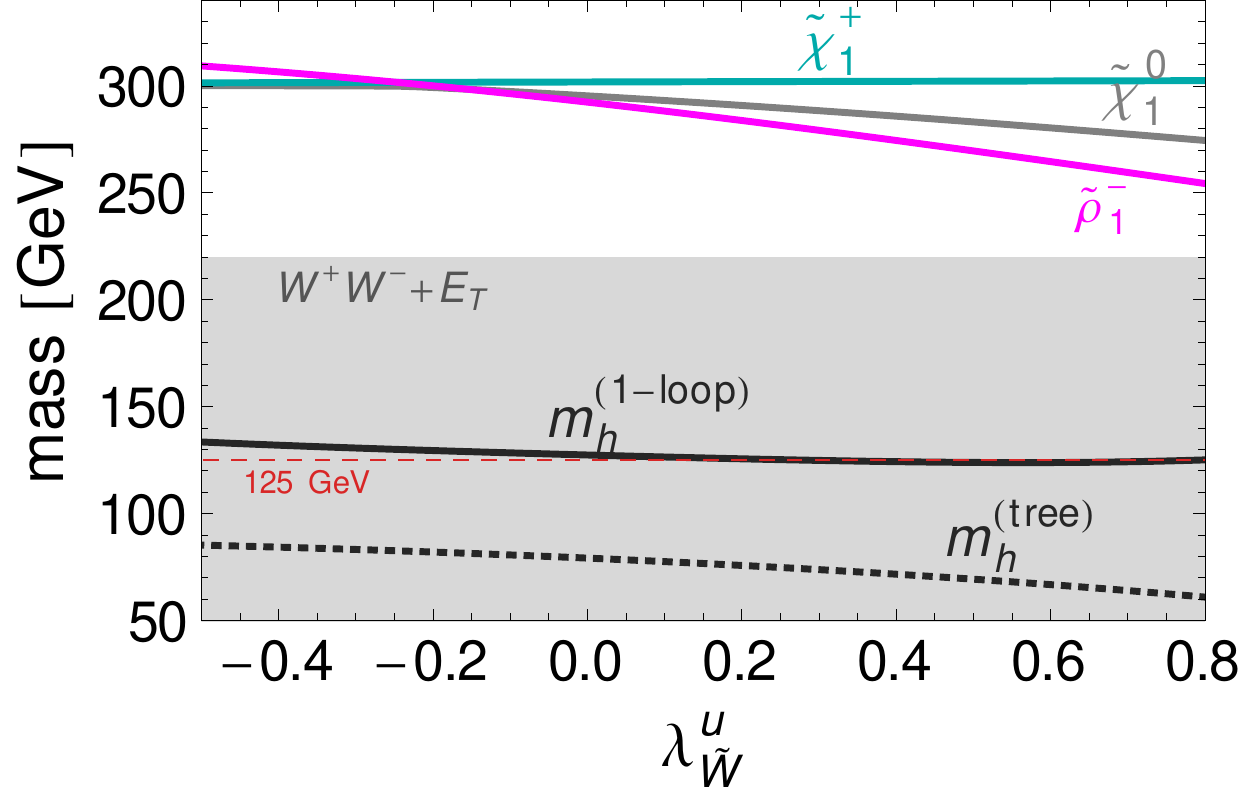}
\caption{ \textbf{Top panel:} Light electroweakino masses (colored lines). The tree- (dotted) and 1-loop (black solid) Higgs mass are shown as a function of $\lambda_{\widetilde{W}}^{u}$, with any other parameterset at the benchmark point $P_{A}$. \textbf{Bottom panel:} Analog to the top panel, but for the benchmark point $P_{B}$.}
\label{fig:mhCharginoNLSP}
\end{figure}

Although one of the original goals of previous works (\cite{Bertuzzo:2014bwa,Diessner:2014ksa}) consisted in maximizing the size of the Higgs mass by first saturating the tree-level piece through the $\lambda$ sign choices (derived from the Higgs mixing entries in Appendix \ref{app:Hmixing})
\begin{equation}
\lambda_{\widetilde{B}}^{u},\lambda_{\widetilde{W}}^{u}<0\text{~~~~~and~~~~~}\lambda_{\widetilde{B}}^{d},\lambda_{\widetilde{W}}^{d}>0~, \label{eq:lambdas}
\end{equation}
these conditions do not automatically guarantee a chargino NLSP regime $\Delta m_{+0}<0$. Back in Sec.\ref{sec:NLSPandLHC}, when describing the mixing matrices (\ref{eq:simpleMchargino}) and (\ref{eq:simpleMneutralino}), we argued that the relative signs between $\lambda$'s are critical to set a $\widetilde{\rho}_{1}^{\pm}$ as the NLSP. We now numerically investigate the extent to which one of the $\lambda^{u}$ can deviate from the Higgs mass sign choice (\ref{eq:lambdas}). To exemplify it, we pick $\lambda_{\widetilde{B},\widetilde{W}}^{d}=+1=-\lambda_{\widetilde{B}}^{u}$, whose signs are exactly as required by the decrease of the Higgs mixing, but set $\lambda_{\widetilde{W}}^{u}=+0.3$, with sign \textit{opposing} the one required to cancel the $H_{u}$-$\phi_{\widetilde{W}}^{0}$ admixture. The specific $\lambda_{\widetilde{W}}^{u}=+0.3$ value will be justified a posteriori from the top panel of Fig. \ref{fig:mhCharginoNLSP}, but the point to remember is that it will ensure a chargino NLSP \textit{and} a correct Higgs mass.

$(m_{h}^{2})_{\text{tree}}$ receives corrections through the one-loop contribution of the CP-even adjoint scalars and the stops to the quartic, as given in Eqs. (\ref{eq:CWpotential}) and (\ref{eq:stopLoop}). Since the latter requires specifying a stop mass, we follow the latest searches \cite{CMS:2017arv,Aaboud:2017ayj} and pick $m_{\widetilde{t}}^{2}=(1.12\tev)^{2}$. Furthermore, in order to follow the no-tachyon condition on the adjoint $B$-terms, these are set to a common value $B_{\text{adj}}=m_{\text{adj}}^{2}/3$ that is numerically safe.

\begin{figure}[h!]
\centering
\includegraphics[scale=0.65]{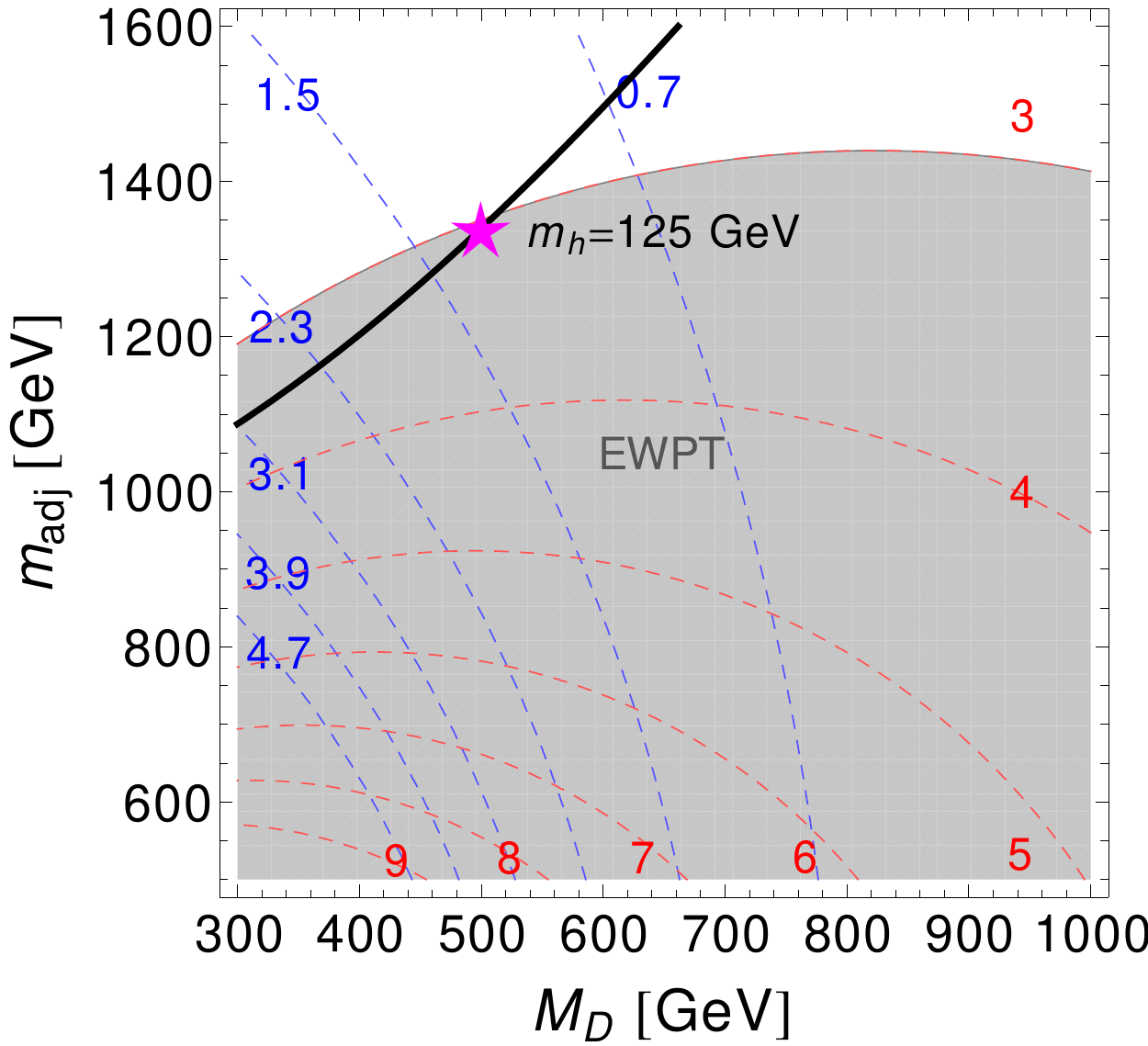}
\caption{The 125 GeV Higgs mass contour (solid) and the tree-level value of the singlet (blue dashed) and triplet (red dashed) adjoint vevs for $\mu=250\gev,~B_{\mu}=400\gev,~\tan \beta=50$, and supersymmetric couplings $\lambda_{\widetilde{B}}^{d}=\lambda_{\widetilde{W}}^{d}=+1,~\lambda_{\widetilde{B}}^{u}=-1$ and $\lambda_{\widetilde{W}}^{u}=+0.3$. Stops are quasi-degenerate and fixed at $1.12 \tev$. In the gray region $v_{\widetilde{W}}$ exceeds the EWPT bound. The star pins down a chargino NLSP benchmark point $P_{A}$.}
\label{fig:mhANDvevs}
\end{figure}

With the current parameter space choices, the singlet and triplet adjoint vevs $v_{\widetilde{B}}$ and $v_{\widetilde{W}}$ are fully specified at each $(M^{D},m_{\text{adj}})$ through their respective minimization conditions, and we display their values as blue ($v_{\widetilde{B}}$) and red ($v_{\widetilde{W}}$) dashed contours in Fig. \ref{fig:mhANDvevs}. By its triplet nature, $v_{\widetilde{W}}$ is subject to EWPT constraints due to its potentially dangerous contribution to the $T$-parameter, and the shaded gray region shows locations where it surpasses the $\approx 3\gev$ bound \cite{Bertuzzo:2014bwa} (see also \cite{Diessner:2014ksa}). Although the vev of the singlet $v_{\widetilde{B}}$ is not limited by EWPT, the fact it shares a similar functional dependence with $v_{\widetilde{W}}$ and has comparable mass parameters would led us to expect similar $O(\text{GeV})$ values. Indeed, this is confirmed by the blue contours in the same plot.

Based on the 125 GeV contours and the $T$-parameter constraint on $v_{\widetilde{W}}$ in Fig.  \ref{fig:mhANDvevs}, a sample point $(M^{D},m_{\text{adj}})=(500,1350)\gev$ is picked (pink star in the same graph) right at the edge of the $v_{\widetilde{W}}$ exclusion. Its purpose is to fix a subset of the parameters governing the EW-inos, which eases their study in the following paragraphs. This is also a necessary step in order to access a sample spectrum in Sec. \ref{sec:spectrum}.

At this stage, the Higgs-adjoint scalars and electroweakino spectrum are completely fixed. For the interested reader, the mixing matrices are collected in Appendix \ref{app:EWmixing}). In summary, our choices of parameters from now on are collectively denoted $P_{A}$,
\begin{align}
P_{A} &:  t_{\beta}=50,~M^{D}=500\gev,~~m_{\text{adj}}=1.35\tev, \notag \label{eq:benchPchar} \\
& \mu=250\gev,~B_{\text{adj}}=\tfrac{1}{3}(1.35\tev)^{2}, \\ \notag 
& \lambda_{\widetilde{B}}^{d}=\lambda_{d}^{\widetilde{W}}=+1,~\lambda_{\widetilde{B}}^{u}=-1,~\lambda_{\widetilde{W}}^{u}=+0.3~.
\end{align}
We must say that $P_{A}$ is merely illustrative. The spectrum and signals for different points can be carried out in a similar way. The relevant behavior to keep in mind for the moving pieces entering in the 1-loop $m_{h}$, but not in the EW-inos (i.e. the adjoint mass parameters) is that all adjoint contributions to $m_{h}$ are proportional to the $\lambda$'s. Keeping these couplings fixed, a larger $m_{\text{adj}}^{2}$ (as well as $B_{\text{adj}}$) decreases the tree-level adjoint term while logarithmically lifting the adjoint loop piece. This contribution competes with the stops correction (\ref{eq:stopLoop}), so larger(smaller) adjoint mass parameters demand lighter(heavier) stops, since if both masses were large one would overshoot the mass of the Higgs.

We proceed to motivate our $\lambda_{\widetilde{W}}^{u}=+0.3$ selection. If we take $P_{A}$ but let $\lambda_{\widetilde{W}}^{u}$ float, the splitting $\Delta m_{+0}$ between the lightest neutralino and lightest chargino would be a function of this coupling. To be able to identify the $\lambda_{\widetilde{W}}^{u}$ size where $\Delta m_{+0}$ flips sign, we look at the variation of the light electroweakino mases as a function of it, with all other quantities set at $P_{A}$. This is precisely what is shown in the upper curves of the top panel of Fig. \ref{fig:mhCharginoNLSP}. Around $\lambda_{\widetilde{W}}^{u}\gtrsim -0.25$ one notices that the $\rho_{1}^{\pm}$ chargino mass goes below the lightest neutralino and becomes the NLSP. For reference, the tree- and loop-level $m_{h}$ are shown, in dotted and dashed lines, on top of  the electroweakino masses. Also shown is the $\widetilde{\rho}_{1}^{\pm}$ mass bound from $W^{+}W^{-}$ in Eq. (\ref{eq:WWxsection}) (gray shade). This chargino NLSP roughly complies with a $125\gev$ Higgs in the $0.1 \lesssim \lambda_{\widetilde{W}}^{u}\lesssim 0.7$ range, backing up our previous selection of $\lambda_{\widetilde{W}}^{u}=0.3$ back in Eq. (\ref{eq:benchPchar}). For reference, at this specific point
\begin{equation}
m_{\widetilde{\rho}_{1}^{\pm}}=225\gev,~~~~~m_{\widetilde{\chi}_{1}^{0}}=236\gev~.
\end{equation}
To briefly compare our $m_{h}$ and EW-ino results against other parameter choices, we show in the bottom panel of Fig. \ref{fig:mhCharginoNLSP} a second benchmark with distinct $M^{D}$, $\mu$ and $m_{\text{adj}}$. We call this benchmark $P_{B}$,
\begin{equation}
P_{B} : M^{D}=600\gev,~~\mu=300\gev,~~m_{\text{adj}}=1.60\tev~. \label{eq:PBbench}
\end{equation}
Clearly, $P_{B}$ displays heavier EW-inos while still accomodating a $\widetilde{\rho}_{1}^{\pm}$ NLSP and $m_{h}$, albeit in a smaller $\lambda_{\widetilde{W}}^{u}$ range. At the same $\lambda_{\widetilde{W}}^{u}=+0.3$, the lightest EW-inos in $P_{B}$ are
\begin{equation}
m_{\widetilde{\rho}_{1}^{\pm}}=275\gev,~~~~~m_{\widetilde{\chi}_{1}^{0}}=287\gev~.
\end{equation}
For the rest of this section, the working $\lambda_{\widetilde{W}}^{u}$ range is translated into a restriction on $m_{\widetilde{\rho}_{1}^{\pm}}$, and until Sec. \ref{sec:spectrum} the benchmark value  $\lambda_{\widetilde{W}}^{u}$ will be used to calculate the scalar and electroweakino spectrum. 

\begin{figure}[h!]
\centering 
\includegraphics[scale=0.65]{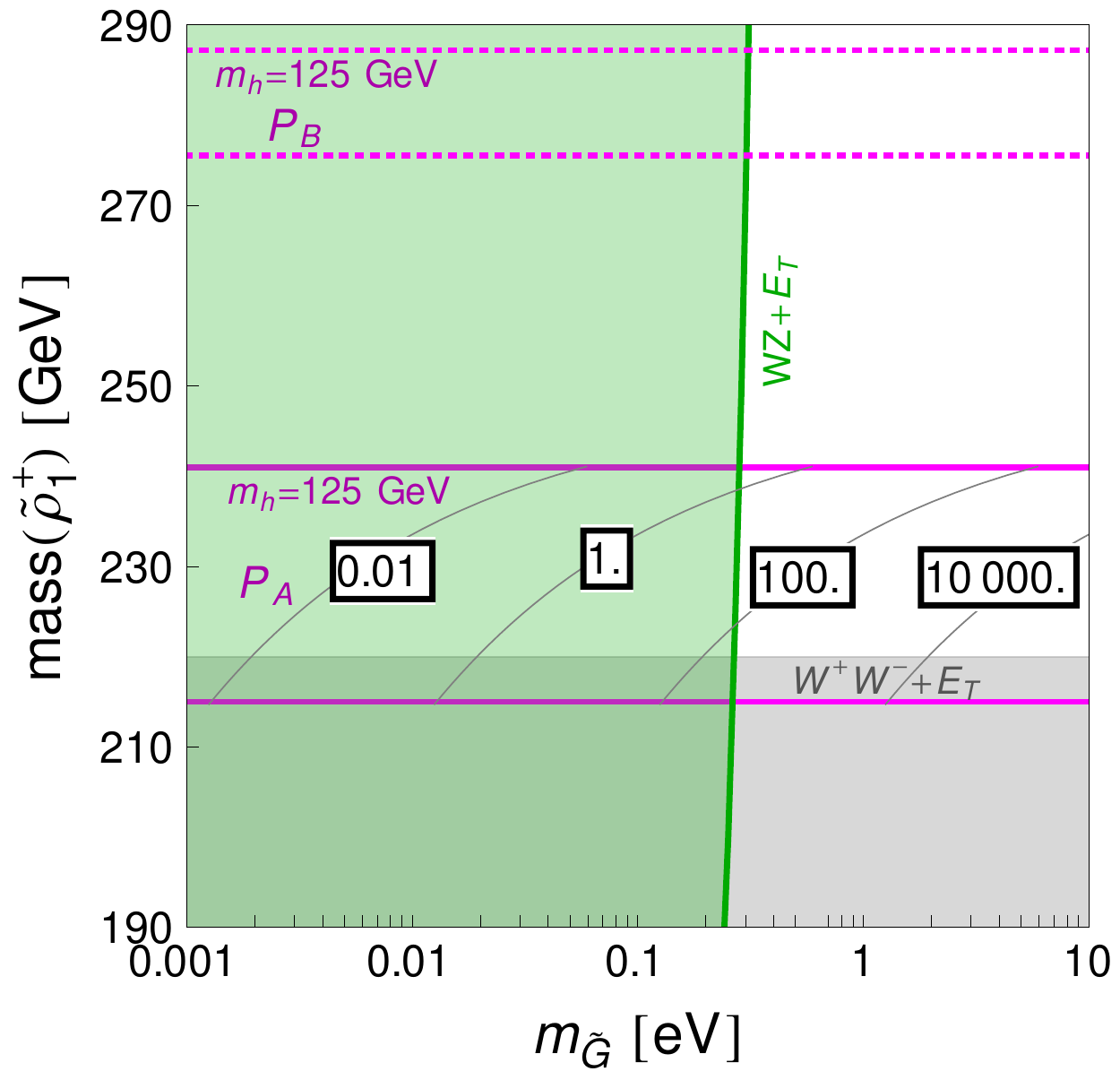}
\includegraphics[scale=0.65]{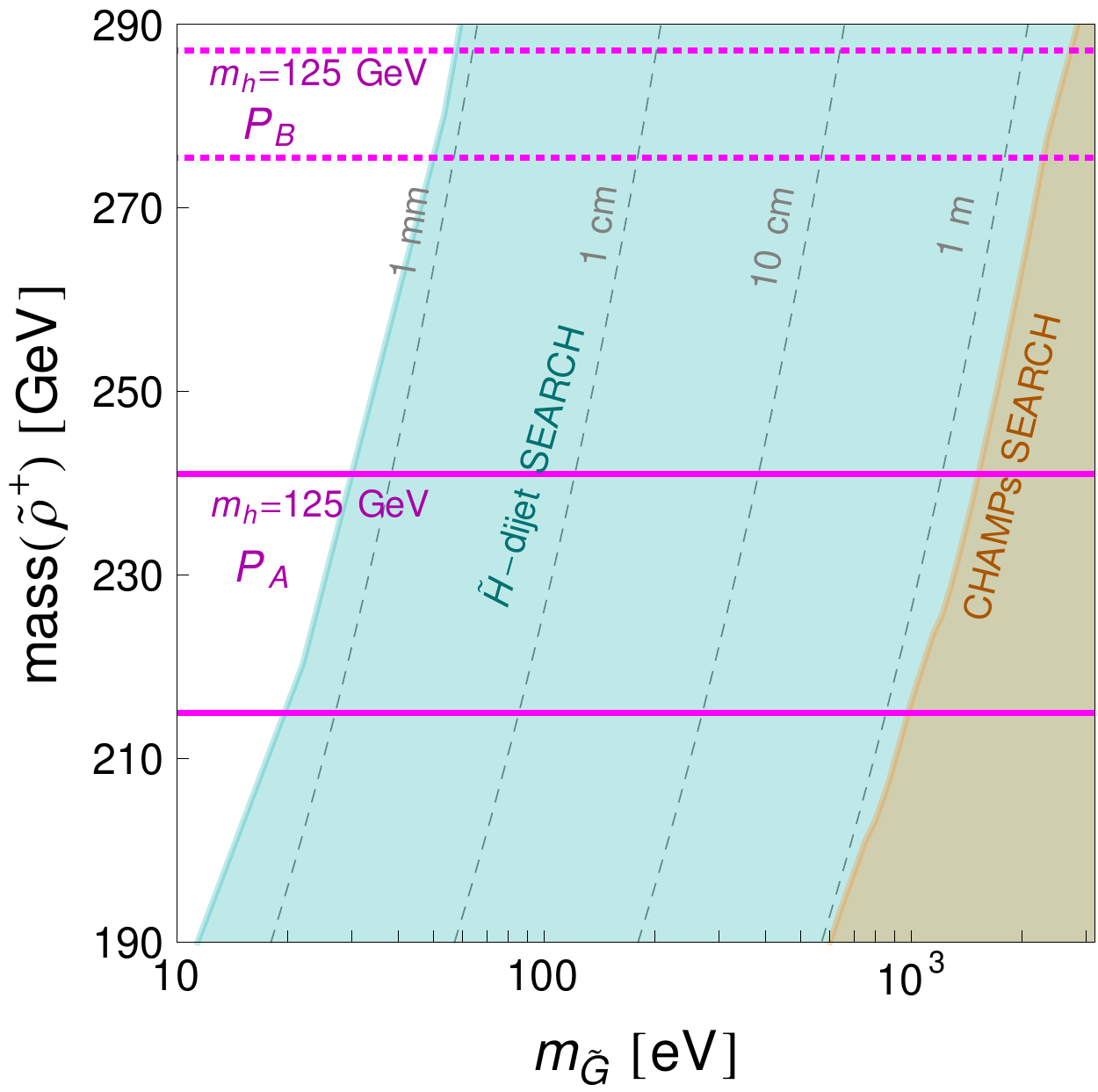}
\caption{ \textbf{Top panel:} Zoomed-in version of the top panel in Fig. \ref{fig:massGmassNLSP} (prompt gravitino regime), with the Higgs mass constraint appearing as the range within the magenta band (solid for $P_{A}$ benchmark, dashed for $P_{B}$). \textbf{Bottom panel:} A zoomed-in version of the bottom panel of Fig. \ref{fig:massGmassNLSP} (longevous gravitino regime), with the Higgs mass constraint superimposed as a magenta band on the chargino NSLP mass axis.}
\label{fig:mGmNLSPhiggs}
\end{figure}

 Having explored the Higgs mass constraint in the MRSSM, we now apply it to the prompt and long-lived regimes of Fig.~\ref{fig:massGmassNLSP}.
 In the top panel of Fig. \ref{fig:mhCharginoNLSP} plot, the $0.1<\lambda_{\widetilde{W}}^{u}<0.7$ range for $m_{h}=125\gev$  shows a one-to-one correspondence with the $215 \gev \lesssim m_{\widetilde{\rho}_{1}}^{\pm}\lesssim 240\gev$ mass interval. We draw this range, corresponding to $P_{A}$, as a band delimited by solid magenta lines in both panels of Fig. \ref{fig:mGmNLSPhiggs}. For the prompt chargino case (top panel), one observes that the $W^{+}W^{-}+\slashed{E}_{T}$ bound falls outside the solid $m_{h}$ band, leaving the $R_{\Gamma}$ green line obtained previously (the $\ell^{+}\ell^{-}+2j+\slashed{E}_{T}$ limit at small $R_{\Gamma}$) and the $m_{h}$ constraint as the only limits across varying gravitino masses. Combining both constraints, there is a $m_{\widetilde{G}}$ lower bound approximately around $0.3\ev$. We remind the reader that at fixed $m_{\widetilde{G}}$, moving vertically within the solid magenta band (i.e. for varying $\lambda_{\widetilde{W}}^{u}$ coupling) implies varying values of the $\Delta m_{+0}$ mass difference, and that this generates distinct $R_{\Gamma}$ curves than in the fixed-$\Delta m_{+0}$ contours of the top panel of Fig.~\ref{fig:massGmassNLSP}. Yet, to motivate the use of the same $R_{\Gamma}$ green line as in the $|\Delta m_{+0}|=10\gev$ case of Fig.~\ref{fig:massGmassNLSP}, we stress that in the $m_{h}=125\gev$ vicinity the value taken by $|\Delta m_{+0}|$ is indeed around $10\gev$ (evident from the top Fig. \ref{fig:mhCharginoNLSP}). When the same solid $m_{h}$ band is superimposed on the bottom panel of Fig.~\ref{fig:massGmassNLSP}, an upper gravitino mass bound is obtained between $20$ and $30 \ev$, where the $m_{\widetilde{G}}$ encounters the higgsino displaced dijet limit in blue. Taking all collider constraints together at $P_{A}$, where $m_{\widetilde{\rho}_{1}^{\pm}}=225\gev$ and $m_{\widetilde{\chi}_{1}^{0}}=236\gev$, the gravitino mass is then restricted between $0.2\ev< m_{\widetilde{G}}<20\ev$. When the previous procedure is redone for the $P_{B}$ benchmark, one gets instead the dashed magenta band, and the upper gravitino mass bound slightly increases to $ m_{\widetilde{G}}<50\ev$.  The $P_{B}$ Higgs mass band is narrower because the working $\lambda_{\widetilde{W}}^{u}$ range in the bottom panel of Fig. \ref{fig:mhCharginoNLSP} is smaller than for $P_{A}$.

Before concluding this section, we comment on discriminating among different higgsino scenarios. Recall that the collider limits shown for the current regime apply to a generic charged higgsino, a NLSP that can also be arranged in a corner of the MSSM parameter space \cite{Kribs:2008hq}. On kinematic grounds, an observable that could help discerning the MRSSM chargino NLSP from the MSSM one is the size of $\Delta m_{+0}$. When this mass difference is large enough, an MSSM scenario is unlikely because even radiatively $|\Delta m_{+0}|\gtrsim 5\gev$ is difficult to arrange. A second handle to help identify the right scenario is the lack of same sign dilepton signals in the Dirac (MRSSM) case. Finally, to distinguish the chargino NLSP from other potential NLSPs, a more comprehensive analysis comparing rates in different channels is required. As one example, slepton NLSPs decay to  same-flavor lepton pairs, while the $W^+W^-$ from chargino NLSPs can decay to all leptons flavor combinations (as well as to jets), so one can look for correlated signals in different lepton flavor bins to differentiate between scenarios.



\section{Full spectrum}
\label{sec:spectrum}
\begin{figure}[h!]
\centering 
\includegraphics[scale=0.65]{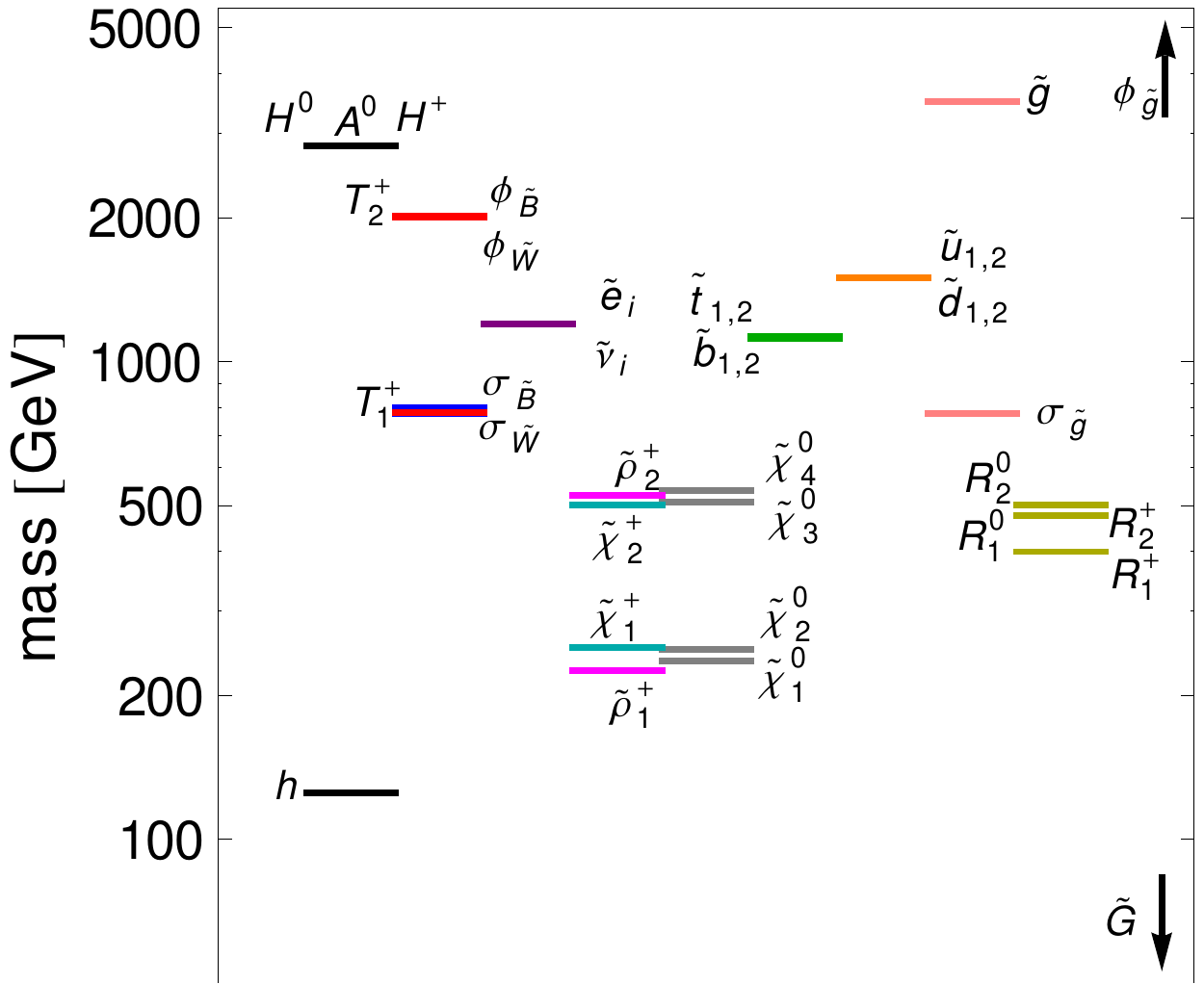}
\includegraphics[scale=0.65]{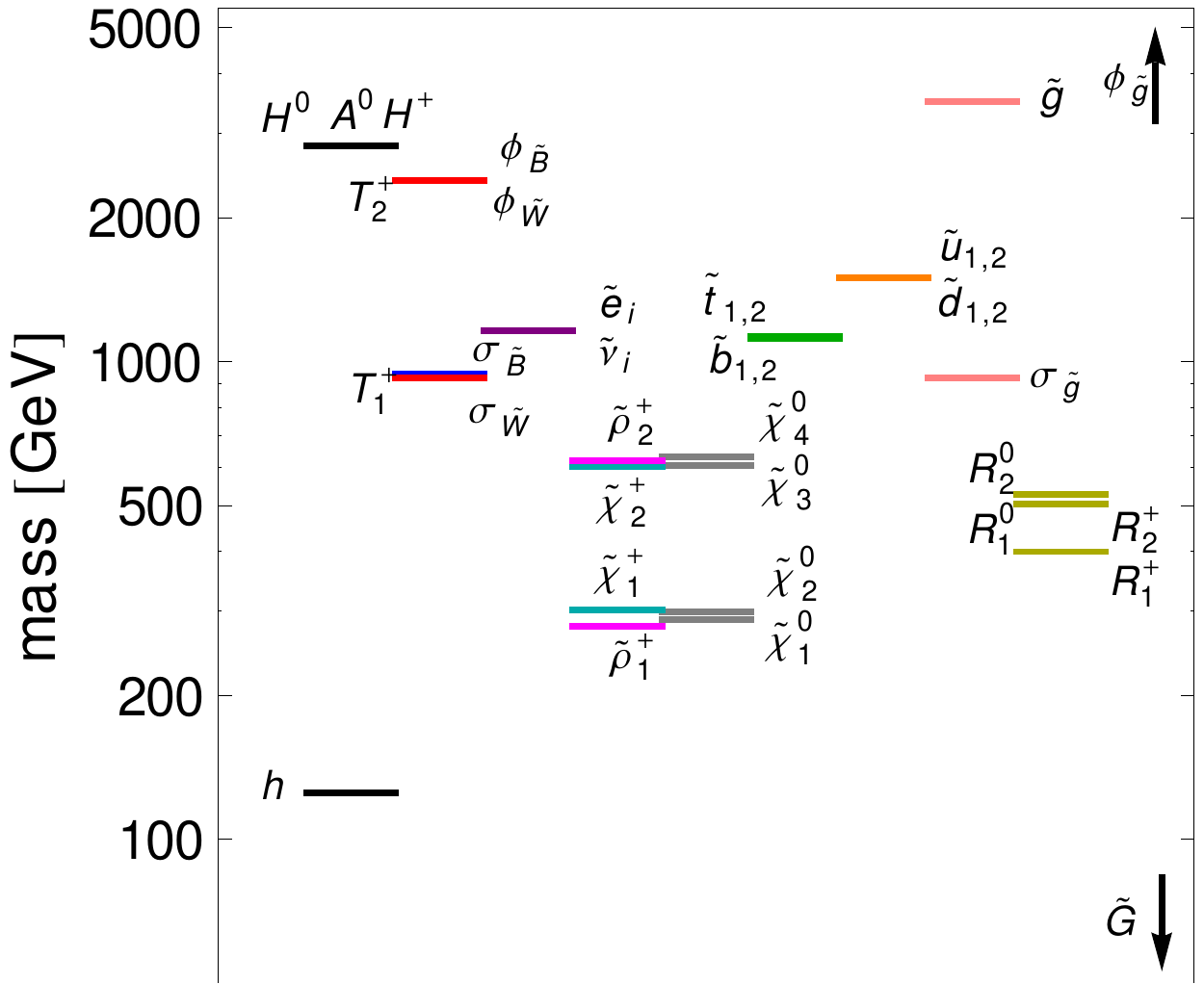}
\caption{\textbf{Top panel:} Spectrum at the benchmark point $P_{A}$. The CP-even sgluon $\phi_{\widetilde{g}}$ sits at $7.2\tev$ and it is not shown. The gravitino, too-light at $m_{\widetilde{G}}\ll 1\gev$, is also outside this range. \textbf{Bottom panel:} The $P_{B}$ benchmark analog to the top panel (see Eq. (\ref{eq:PBbench}))}
\label{fig:spectrumPbench}
\end{figure}

With the purpose of offering a complete low-energy model, we now present the remaining parts of the spectrum at $P_{A}, P_{B}$. The physical masses are sketched in Fig. \ref{fig:spectrumPbench} except for $\phi_{\widetilde{g}}$ (too heavy) and $\widetilde{G}$ (too light). All of them have been verified with \texttt{SPheno 3.3} via the \texttt{SARAH 4.8} implementation of the MRSSM \cite{Porod:2003um,Staub:2008uz}. To continue, some relevant comments are included for each sector.

\textit{Higgs sector:} Among the CP-even scalars, the SM-like Higgs $h$ is mostly up-type, the next two heavier neutral states $\phi_{\widetilde{B}},~\phi_{\widetilde{W}}$ are mostly-singlet and mostly-triplet mixtures; the heaviest $H^{0}$ is dominantly down-type. Besides the $Z$ Goldstone, there are 3 pseudoscalars: the heaviest one $A^{0}$ dominated by $H_{d}^{0}$, and the other two are adjoint-mixtures $\sigma_{\widetilde{B},\widetilde{W}}$. Similarly, the four charged scalars mix into a $W^{\pm}$ Goldstone, a heavy, charged down-type Higgs $H^{+}$, and two charged $T_{1,2}^{+}$. Up to a small mixing with the Higgs doublets, the physical adjoint masses are mainly set by the size of $m_{\phi_{a}}^{2}$ and $m_{\sigma_{a}}^{2}$ (themselves larger than $M^{D}$ and the $gv$-sized pieces) in Eq. (\ref{eq:scalarAdjSoft}). No $R=+2$ scalar enters in this category because their neutral and charged states do not mix with the up/down doublets or with the adjoints.

\textit{Electroweakinos:} Denoting the fermion of the $R_{u,d}$ doublets by \textit{R}-higgsinos, the four neutralinos $\widetilde{\chi}_{i}^{0}$ are divided into two $R$-higgsino-higgsino mixtures, and two gaugino-adjoint fermion combinations. One can show (see Appendix \ref{app:EWmixing}) that the charginos are split into two disconnected sets ($\widetilde{\chi}$-type and $\widetilde{\rho}$-type) due to the different electric and $R$-charges. Still, both of these sets display a light $H$-$R$ higgsino and a heavy electroweak Dirac gaugino.

\textit{Gluino and sgluons:} The supersoft origin of $M_{\widetilde{g}}$ permits a several-TeV heavy gluino without introducing large fine-tuning. With this in mind, we pick $M_{\widetilde{g}}=3.5\tev$, which also becomes the physical mass of $\widetilde{g}$ because it is the only colored fermion octet. 

Similarly, being the only scalar color octets, the two sgluon states are already physical CP eigenstates with masses determined by
\begin{align}
     m_{\phi_{\tilde{g}}}^{2} &= m_{\text{adj}}^{2}+2B_{\text{adj}}+4(M_{\widetilde{g}}^{D})^{2} \notag \\
m_{\sigma_{\tilde{g}}}^{2} &= m_{\text{adj}}^{2}-2B_{\text{adj}}~. \label{eq:scalarAdjSoft}
\end{align}
For the CP-even state, in the upper equation in (\ref{eq:scalarAdjSoft}), the third contribution originates from the supersoft operator (\ref{eq:CSO}). Hence, the several-TeV gluino mass causes $\phi_{\widetilde{g}}^{2}$ be considerably heavier ($\approx 7.21\tev$, not shown in the figure) than its pseudoscalar counterpart, which stays at near $780\gev$.

\textit{$R$-scalars:} The $R=+2$ scalars do not mix with the Higgs states or the adjoints. The neutral $R_{u,d}^{0}$ have each a $m_{R}^{2}$ soft mass and become states $R_{1,2}^{0}$, which up to a $O(\lambda^{2}v^{2})$ mixing amount are effectively mass eigenstates. Regarding the charged components, the $R_{u}^{-}$ and $R_{d}^{+}$ mixing is prevented by the $R$-symmetry at the superpotential level, and these become physical states $R_{1,2}^{+}$. The $R$-scalar spectrum is then set by its soft mass only, and for definiteness $m_{R}^{2}=(400\gev)^{2}$ is picked.

\textit{Sfermions:} The sfermions count with two sources for their soft masses, the ($D$-breaking) finite log from (\ref{eq:finiteLog}), and the $F$-breaking piece (\ref{eq:softNonHolo}). Clearly, in the strictly supersoft limit, the finite log stablishes a fixed hierarchy $\sim(4\pi/\alpha_{a})^{1/2}$ between the gaugino and sfermion masses for each $M_{a}^{D}$ value.  Among squarks, the finite log is well approximated by the color term alone due to the $\alpha_{s}$ size, and as an example, at $M_{\widetilde{g}}=3.5\tev$ we have $m_{\widetilde{q}}^{2}=(945\gev)^{2}$. On the other hand, current mass bounds on 1st and 2nd generation squarks ($\widetilde{q}_{1,2}$) place them at no less than $1.5\tev$ \cite{ATLAS-CONF-2017-022}, thus these can be made sit right at the bound for a $F$-piece not smaller than $(m_{\text{soft}^{2}})_{F}=(1200\gev)^{2}$. The corresponding bounds for stops/sbottoms are weaker, laying around $1.12\tev$, a value that is reproduced by a  $F$-breaking mass about half the size ($\approx 600\gev$) of the one used for the $\widetilde{q}_{1,2}$.

Despite the similar origin between slepton and squark mass contributions, there are critical numerical differences. In the absence of $F$-breaking, the proportionality of slepton masses to the electroweak $\alpha_{a=1,2}$ implies smaller finite logs compared to the squarks, even if all three gaugino masses were comparable. However, too-light sleptons can be troublesome, and require raising $M^{D}$ accordingly. But once additional $F$-breaking is turned on, the need of increasing $M^{D}$ to raise $m_{\widetilde{\ell}}^{2}$ is partially removed: in principle we can take advantage of this effect to enforce the chargino as a NLSP by providing an $F$-term piece just large enough to avoid any slepton becoming the NLSP. Yet, these $F$-terms are chosen even heavier so as to make sleptons heavier than the rest of the electroweakinos and effectively removing them from the NNLSP decays\footnote{Intermediate slepton mass values between the $\widetilde{\chi}_{2}^{0}$ and $\widetilde{\rho}_{1}^{\pm}$ causes $\widetilde{\chi}_{2}^{0}$ to decay through $\widetilde{\ell}$ instead.}. For simplicity, we adopt the very same $(m_{\text{soft}}^{2})_{F}=(1200\gev)^{2}$ as for the 1st and 2nd generation squarks. Hence, all sfermion soft masses are built from a common $F$-term piece and a $\alpha_{a}$-dependent supersoft part, $m_{\widetilde{f}}^{2}=(m_{\widetilde{f}}^{2})_{D}+(m_{\text{soft}}^{2})_{F}$.

We finish the current section by drawing in the bottom panel of Fig. \ref{fig:spectrumPbench} the corresponding spectrum at the benchmark $P_{B}$ in Eq. (\ref{eq:PBbench}). In this case the EW-inos and the mostly-adjoint scalars are heavier, a consequence of having chosen larger $\mu$, $M^{D}$ and $m_{\text{adj}}$ values. Likewise, larger contributions to the fine-tuning are expected for $P_{B}$.



\section{Fine-tuning estimate}
\label{sec:FT}

To complement our numerical discussion,  we present a calculation of the FT. As already stated, the presence of new scalar sector mass parameters and mixing with extra states modifies the Higgs minimization condition w.r.t. the MSSM case. The purpose of this section is to identify numerically the level of FT around the parameter benchmark adopted in the previous section.

The different contributions to the FT, their relative sizes and their consequences for the interplay between scalar adjoint, Dirac gaugino and stop masses have been studied in detail in Ref.\cite{Bertuzzo:2014bwa}, following the FT measure defined in \cite{Gherghetta:2012gb}. Let us quote it here: the fine-tuning in the electroweak vev (or equivalently, in $m_{Z}^{2}$)  is quantified through $\Delta_{v}\equiv   \text{max}_{i}\{\Delta_{i}\}$,
\begin{equation}
 \text{max}_{i}\{\Delta_{i}\}=\text{max}_{i}\biggl| \sum_{j}\dfrac{\xi_{i}(\Lambda_{\text{mess}})}{m_{Z}^{2}}\dfrac{m_{Z}^{2}}{d \xi_{j}(m_{\text{soft}})}\dfrac{d\xi_{j}(m_{\text{soft}})}{d\xi_{i}(\Lambda_{\text{mess}})} \biggr|  \label{eq:deltaFT}
\end{equation}
where $i$ runs over the input parameters that fix the value of $m_{Z}^{2}$\footnote{The Higgs mass FT measure $\Delta_{m_{h}}$ takes an analogous form.}. For the present model the parameters are $\xi_{i}=m_{\text{adj}}^{2},~m_{R}^{2},~m_{\widetilde{t}}^{2},~\mu,~M^{D},~B_{\text{adj}},~\lambda_{\widetilde{B}}^{u}$ and $\lambda_{\widetilde{W}}^{u}$, where for simplicity we've already set common adjoint, gaugino and higgsino masses. For the stops, the inherent absence of the $A_{t}$ trilinear and  MSSM $\mu$-term implies vanishing LR mixing. It follows then that $m_{Q_{3},u_{3}}^{2}$ become the physical masses, mostly set by the finite log in Eq. (\ref{eq:finiteLog}) and differing only by the subdominant wino piece acquired by $m_{Q_{3}}$. In our notation, $m_{\widetilde{t}}$ will refer to their geometrical average. Only the $u$-type $\lambda$'s take part in the $\xi_{i}$ list because we have restricted ourselves to the large $\tan\beta$ limit (refer to Sec. \ref{sec:mHiggs}). Clearly, the adjoint vevs $v_{\text{adj}}$ are not included because they are traded by combinations of the $\xi_{i}$ via their (coupled) minimization equations.

The analysis in Ref.\cite{Bertuzzo:2014bwa} showed that the dominant $\Delta_{v}$ contributions are $\Delta_{\mu}$ (tree level), $\Delta_{R_{u}}$, $\Delta_{\widetilde{t}}$, and $\Delta_{\widetilde{W},\widetilde{B}}$ (loop-level). As such, the FT mainly depends on the respective mass scales $\mu, m_{R}, m_{\widetilde{t}}$ and $m_{\text{adj}}$, plus the messenger cutoff appearing as $\log{(\Lambda_{\text{mess}}}/m_{\text{soft})}$ in the one-loop pieces (with $m_{\text{soft}}$ set at the stop mass). Contributions of $M^{D}$ to $\Delta_{v}$ are, however, subleading. The Appendix \ref{app:FT} collects the explicit dependence of the largest $\Delta_{i}$'s. The cutoff $\Lambda_{\text{mess}}$ must be compatible with low-energy mediation but it is not fixed by our benchmark, nor are the $F$- and $D$-breaking spurions. The reason is that so far we have been working directly with the ratios $\langle D\rangle/\Lambda_{\text{mess}}$ (Dirac gaugino masses) and $\langle F\rangle/\Lambda_{\text{mess}},~\langle D\rangle/\Lambda_{\text{mess}}$ (for sfermion and adjoint soft masses). In the next paragraphs we look at the behavior of the FT as a function of $\Lambda_{\text{mess}}$.

The FT is now estimated around $P_{A}$ (for $P_{B}$ will be somewhat larger since the whole spectrum is a bit heavier) in Eq. (\ref{eq:benchPchar}). Given the various scales involved ($\mu,~m_{\text{adj}},~m_{\tilde{t}},~\Lambda_{\text{mess}}$), extracting useful FT information requires us to pick appropriate 2D planes for these masses. We opt  to work in the $(\Lambda_{\text{mess}},m_{\text{adj}})$ plane at constant (i.e. benchmark) $\mu$ and $m_{\tilde{t}}$, as all FT pieces (except for $\Delta_{\mu}$) depend on the cutoff, and two of them ($\Delta_{\widetilde{B},\widetilde{W}}$) depend on $m_{\text{adj}}$. Specifically, we vary $m_{\text{adj}}$ might be varied between $1.2-1.6\tev$ (safe from EWPT in Fig. \ref{fig:mhANDvevs}) while the messenger scale is varied within the low-energy mediation range $10^{4}-10^{14}\text{ GeV}$ of minimal GMSB \cite{Giudice:1998bp} (as explained earlier in Sec. \ref{sec:intro})\footnote{Keeping $m_{h}=125\gev$ in this $m_{\text{adj}}$ interval requires a deviation from the benchmark $M^{D}$ as large as 200 GeV, from Fig. \ref{fig:mhANDvevs}. Yet, $M^{D}$ adds to the FT subdominantly.}.

\begin{figure}[h!]
\centering 
\includegraphics[scale=0.65]{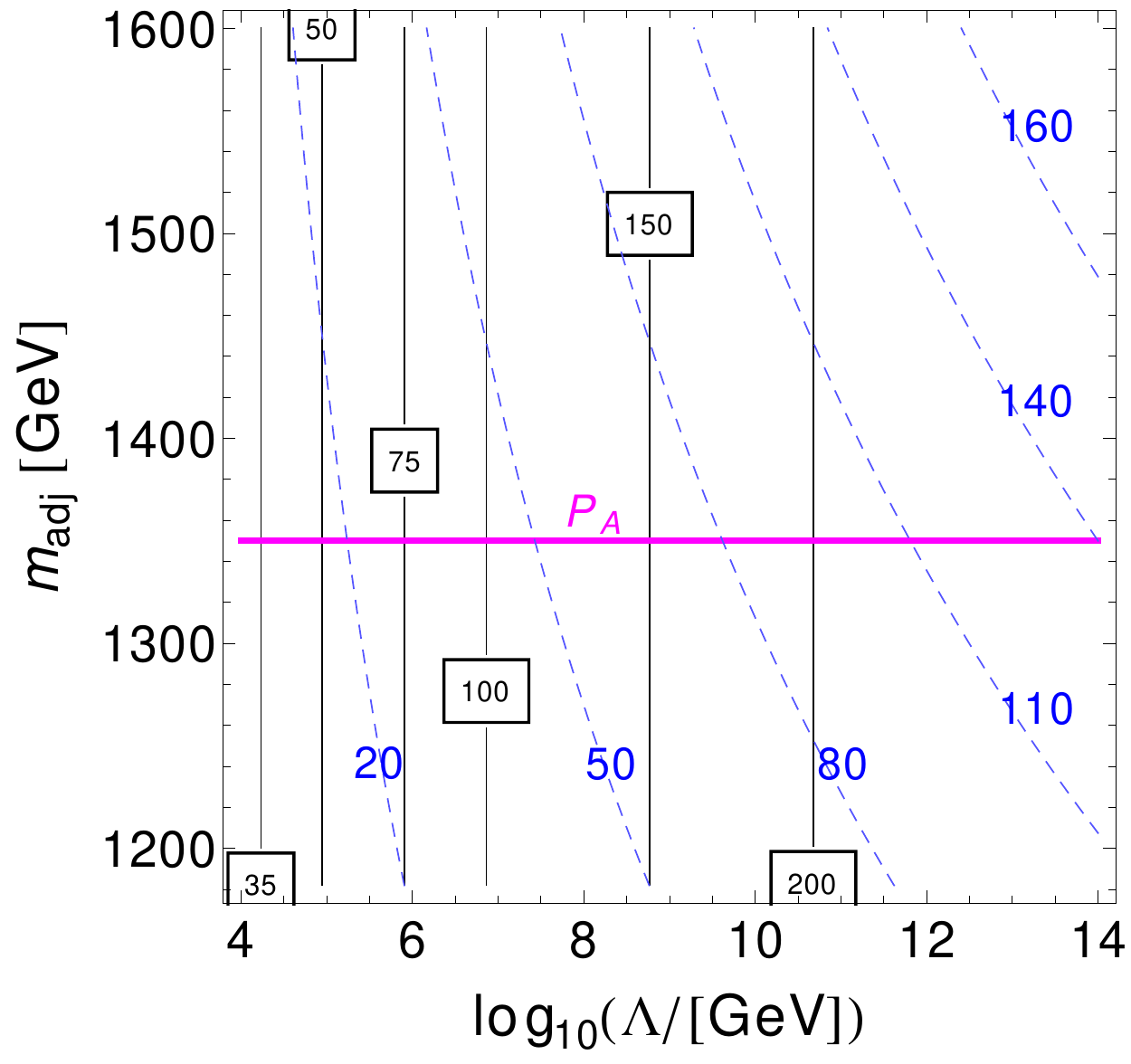}
\caption{Measure $\Delta_{v}$ of fine-tuning (vertical lines) and subdominant contribution $\Delta_{\widetilde{B}}$ from the singlet adjoint (blue curves) as a function of the messenger scale and the (common) adjoint soft mass. The Higgs mass benchmark value used for $m_{\text{adj}}$ is indicated by the magenta line (point $P_{A}$ in Eq. (\ref{eq:benchPchar})).}
\label{fig:FTcontours}
\end{figure}

The overall FT measure is portrayed in Fig. \ref{fig:FTcontours} as black solid lines. Numerically, $\Delta_{v}$ is dominated by the stop contribution; as it does not depend on $m_{\text{adj}}$, the contours are vertical. Also shown are the $\Delta_{\widetilde{B}}$ piece (blue dashed) and the $m_{\text{adj}}$ value at $P_{A}$ (magenta line). As it depends only on $\mu$ (set at 250 GeV), $\Delta_{\mu}\approx 30$ at every point in the plane.  If we want to acheive better than percent-level fine-tuning, we can see that  $\Lambda_{\text{mess}}$ must be less than $\sim 10^{7}\gev$.

How do these statements change for other $\mu$ or $m_{\tilde{t}}$ choices? Larger (smaller) $\mu$ values give a larger (smaller) $\Delta_{\mu}$, again uniform over the whole plane. On the other hand, increasing stop masses still result in vertical $\Delta_{\widetilde{t}}$ contours, but, by the 125 Higgs condition, lighter $m_{\text{adj}}$ are demanded. Then, the pink line (representing a benchmark with correct Higgs mass) would sit lower. In this case $\Delta_{\widetilde{t}}$ would still dominate the FT, and the plot would look similar but with higher $\Delta_{v}$ label values. 

Before finishing the current section and concluding, we compare the relative sizes between $m_{\text{soft}}$, the messenger scale, and the supersymmetry breaking vevs\footnote{ Again, with $F$ and the $F'$ of Eq.(\ref{eq:mass32}) developing comparable vevs.} $\langle F\rangle,~\langle D\rangle,$ at $P_{A}$. This is all done at a particular gravitino mass near the lower limit in Fig. \ref{fig:mGmNLSPhiggs}, say $1\ev$, for illustration. In doing so we remind ourselves of the mass relations $M^{D}\sim \langle D\rangle/\Lambda_{\text{mess}}$ and $m_{\widetilde{G}}\sim \langle F\rangle/M_{P}^{*}$, assuming the $N_{\text{mess}}=1$ gauge-mediation relation $m_{\text{soft}}\sim \langle F\rangle/\Lambda_{\text{mess}}$. With $(m_{\text{soft}}^{2})_{F}=(1200\gev)^{2}$ (the common $F$-piece for sfermions and the squarks of the first two generations), and replacing $\langle F\rangle$ by $m_{\widetilde{G}}M_{P}^{*}$, it holds that
\begin{equation*}
\Lambda_{\text{mess}}\sim \dfrac{m_{\widetilde{G}}[\text{GeV}](10^{18}\gev)}{1.2\times 10^3 \gev}\sim \dfrac{m_{\widetilde{G}}[\text{GeV}]}{1.2\times10^{-15}}~.
\end{equation*}
At a gravitino mass of $1\ev$, the equation above gives us $\Lambda_{\text{mess}}\sim 8\times 10^{5}\gev$, near the lower end of the low-mediation range. At that messenger scale, and the benchmark gaugino mass $M^{D}=500\gev$,
\begin{equation*}
\sqrt{\langle F\rangle}\approx 3\times 10^{4}\gev,~~~~~\sqrt{\langle D\rangle}\approx 2\times 10^{4}\gev~,
\end{equation*}
indicating spurions of comparable size\footnote{ These sizes or course will be affected to some extent when going to $N_{\text{mess}}>1$ and with distinct messenger scales for $\langle F\rangle$ and $\langle D\rangle$. }.


\section{Conclusions}
\label{sec:conclusion}

Model building with Dirac gauginos has led to phenomenological improvements over minimal supersymmetric scenarios. This has been exemplified in the literature by the studies of their effects in flavor, fine-tuning, and the relaxation of LHC bounds on color sparticles. More recently, a renewed wave of interest has taken Dirac gauginos and its $R$-symmetric setups in additional directions, for example by extending the soft terms to include non-supersoft operators \cite{Martin:2015eca}, taking advantage of the enlarged field content to propose dark matter candidates \cite{Goodsell:2015ura}, or studying the collider prospects of the $SU(3)_{c}$ scalar adjoints (sgluons) \cite{Abel:2013kha,Kotlarski:2016zhv}. At the same time, some studies have deepened into the UV aspects of Dirac gauginos, revealing that there are persistent issues to be taken care of, for example the analog of the GMSB $\mu-B_{\mu}$ problem between the Dirac gaugino masses and their corresponding adjoint $B_{a}$-terms \cite{Csaki:2013fla}. Some ideas on this direction can be found in \cite{Nelson:2015cea, Alves:2015kia}.

The present work attempts to complement these categories of models by providing a study of the compatibility between the chargino NLSP regime under LHC-13 constraints and $m_{h}=125\gev$ in the MRSSM (in its incarnation with $R$-symmetric Higgs sector). To do so, we have first shown what the collider limits look like across the chargino-gravitino plane. Next, by reaching the Higgs mass at one loop with the help of $\sim500\gev$ electroweak gaugino Dirac masses and $>1\tev$ adjoint scalar soft masses (together with the stop quantum correction), we delimited a range where the $m_{\widetilde{\chi}_{1}^{0}}>m_{\widetilde{\rho}_{1}^{\pm}}$ mass ordering is maintained under a variation of the (supersymmetric) Higgs-adjoint couplings.  Then, after translating this $m_{h}=125\gev$ requirement back to the mentioned plane, the collider-safe region of the prompt regime is found to be mildly-dependent on the (lightest) chargino-neutralino splitting at sub-eV $m_{\widetilde{G}}$. On the other hand, gravitino masses falling on the $O(10)\ev$ ballpark are subject to the displaced dijet bound. To anchor some numbers, in our sample benchmark with chargino NLSP and the observed Higgs mass value, the prompt and displaced dijet collider searches together safely place gravitinos between $0.2\ev \lesssim m_{\widetilde{G}} \lesssim 20\ev$. The neutralino NNLSP and chargino NLSP respectively sit at $236$ and $225 \gev$.

As presented here, it is clear that at the numerically analyzed points the extra requirement of a chargino NLSP scenario represents a tradeoff of some fine-tuning. To understand this statement, one should remember that this regime imposes conditions on the size and sign of the supersymmetric Higgs-adjoint couplings (the $\lambda$ parameters). These conditions, however, do not strictly align with those that fully saturate the tree-level Higgs mass piece. Thus, not having all the $\lambda$ couplings close enough to 1 has a reducing effect on the adjoints contribution to the Higgs mass. In turn, this demands a larger $m_{h}$ lift from the stops, which at TeV-sized masses become the dominant source of FT (at fixed $\mu$). Another unavoidable feature of our model is that having analyzed a low-energy scenario below the messenger mass scale, our choices of relative sizes between the mass parameters of the Dirac gauginos and adjoints do not necessarily comply with UV considerations (i.e. the $B_{\text{adj}}\sim 16\pi^{2}M^{D}$ relation predicted in supersoft scenarios with no messenger mixing).  These are otherwise evaded by adding more messenger pairs and/or introducing direct interactions with SM superfields \cite{Csaki:2013fla}, but these considerations fall out of the scope of our work.

Experimental hints pointing towards the existence of scenarios with simultaneous $R$-symmetric chargino NLSP and numerically correct Higgs are nontrivial to identify. While the LHC sets constraints on the chargino NLSP alone based on final states with $E_{\slashed{T}}$ and dilepton (with and without jets) or as macroscopic displaced dijet tracks, experimental access to the NNLSP is auxiliary in telling apart setups with $R$-symmetry from those without it. To do that, one strategy is to look at the size of the NNLSP-NLSP splitting: $\Delta m_{+0}$ of a few tens of GeV are ruled out for MSSM-like models at large $\tan\beta$ but realizable in the $R$-symmetric case. Meanwhile, the Higgs mass requirement selects a preferred range for $\widetilde{\rho}_{1}^{\pm}$, yet we have shown that this NLSP range is not unique but dependent on parameters that also determine Higgs-adjoint mixtures and stop masses. Therefore, extra discerning power is gained by looking at these other states too. One final experimental consideration is the fact that in a model where gauginos are Dirac there are no processes leading to same sign dileptons, so one can also use that fact as a discriminator between models preserving the R-symmetry and models like the MSSM.

In conclusion, the present study shows the current status, in terms of the corresponding parameter space, of the chargino NLSP that decays into a gravitino. In a more general sense, it exemplifies the kind of regions where a full, realistic $R$-symmetric Dirac gaugino spectrum is pushed in order to survive up-to-date bounds. This shows once again how non-minimal models can still improve naturalness with respect to minimal supersymmetric setups but cannot be completely devoid of fine-tuning.



\section*{Ackwnowledgements}
\label{sec:ackw}

We thank Jared Evans for valuable suggestions regarding higgsino searches with displaced dijets. This work was partially supported by the National Science Foundation under Grants No. PHY-1417118 and No. PHY-1520966. We 


\appendix
\section{Electroweakino and pseudoscalar mass mixing}
\label{app:EWmixing}

The neutralino mixing matrix $\mathcal{M}_{\widetilde{\chi}^{0}}$, in the basis $(\widetilde{B},\widetilde{W}^{0},\widetilde{R}_{d}^{0},\widetilde{R}_{u}^{0})\times(\psi_{\widetilde{B}},\psi_{\widetilde{W}},\widetilde{H}_{d}^{0},\widetilde{H}_{u}^{0})$, looks like
\begin{equation}
\left(\begin{array}{cccc}
 M^{D}+O(gv_{a}) & 0 & -\tfrac{g'c_{\beta}v}{2} & \tfrac{g's_{\beta}v}{2} \\
0 & M^{D} & \tfrac{gc_{\beta}v}{2} & -\tfrac{gs_{\beta}v}{2} \\
\tfrac{\lambda^{d}c_{\beta}v}{\sqrt{2}} & -\tfrac{\lambda^{d}c_{\beta}v}{2} & \mu+O(\lambda v_{a}) & 0 \\
\tfrac{\lambda^{u}s_{\beta}v}{\sqrt{2}} & -\tfrac{\lambda^{u}s_{\beta}v}{2} & 0 & \mu+O(\lambda v_{a})
\end{array}\right)
\end{equation}
The $R=Q$ charginos, named $\widetilde{\chi}_{1,2}^{\pm}$, have a mixing matrix
\begin{equation}
\mathcal{M}_{\widetilde{\chi}^{\pm}}=
\left(\begin{array}{cc}
M^{D}+O(gv_{a}) & \tfrac{\lambda^{d}c_{\beta}v}{\sqrt{2}} \\
\tfrac{gc_{\beta}v}{\sqrt{2}} & -\mu-O(\lambda v_{a})
\end{array}\right)~,
\end{equation}
in the $(\psi_{\widetilde{W}}^{-},\widetilde{H}_{d}^{-})\times(\widetilde{W}^{+},\widetilde{R}_{d}^{+})$ basis. The $R=-Q$ charginos, $\widetilde{\rho}_{1,2}^{\pm}$, mix according to
\begin{equation}
\mathcal{M}_{\widetilde{\rho}^{\pm}}=
\left(\begin{array}{cc}
M^{D}-O(gv_{a}) & \tfrac{gs_{\beta}v}{\sqrt{2}} \\
\tfrac{\lambda^{u} s_{\beta}v}{\sqrt{2}} & -\mu-O(\lambda v_{a})
\end{array}\right)
\end{equation}
in the $(\widetilde{W}^{-},\widetilde{R}_{u}^{-})\times(\psi_{\widetilde{W}}^{+},\widetilde{H}_{u}^{+})$ basis. Again, the $\sqrt{2}$ in some entries is due to different $\lambda_{\tilde{W}}$ normalization with respect to \cite{Bertuzzo:2014bwa}. Within our numerical analysis at $P_{A}$, the $\widetilde{\chi}_{1}^{\pm}$ is a mostly-higgsino NLSP and $\widetilde{\chi}_{1}^{0}$ is the NNLSP.

The pseudoscalar mixing matrix in the $\bigl( \text{Im}(H_{d}^{0}), \text{Im}(H_{u}^{0}),\sigma_{\widetilde{B}},\sigma_{\widetilde{W}} \bigr)$ basis has a form
\begin{equation}
\mathcal{M}_{\text{CP-odd}}^{2}=
\left(\begin{array}{cc}
\mathcal{M}_{\text{MSSM}} & \boldsymbol{0} \\
\boldsymbol{0} & \mathcal{M}_{\text{singlet-triplet}}  
\end{array}\right)
\end{equation}
where $\mathcal{M}_{\text{MSSM}}$ is the usual MSSM pseudoscalar block and $\mathcal{M}_{\text{singlet-triplet}}$ is
\begin{equation}
\left(\begin{array}{cc}
m_{\text{adj}}^{2}-2B_{\text{singlet}}+O(\lambda^{2}v^{2}) & O(\lambda^{2}v^{2}) \\
O(\lambda^{2}v^{2})                                             & m_{\text{adj}}^{2}-2B_{\text{triplet}}+O(\lambda^{2}v^{2})
\end{array}\right)~.
\end{equation} 

\section{Electroweakino-to-gravitino partial widths}
\label{app:gravitinoWIDTH}
 
In the effective limit of the gravitino as the Goldstino, the partial width of an electroweakino $\widetilde{\chi}$ into a SM gauge boson $V$ and a gravitino is
\begin{equation}
\Gamma(\widetilde{\chi}\to V\widetilde{G}) =\kappa_{\widetilde{G}V}\dfrac{m_{\widetilde{\chi}}^{5}}{96\pi M_{P}^{*2}m_{\widetilde{G}}^{2}}\left[ 1-\dfrac{M_{V}^{2}}{M_{\widetilde{\chi}}^{2}} \right]^{4}  \label{eq:widthChi}~.
\end{equation}
At large $\tan \beta$, the $O(1)$ coefficients in front are  $\kappa_{V\widetilde{G}}=1,~1,~(M_{W}/M^{D})^{2}s_{W}^{2}$ respectively for  $V=W,Z,\gamma$. The $R_{\Gamma}$ ratio of Eq.(\ref{eq:Rgamma}) is given by \cite{Kribs:2008hq}
\begin{equation}
R_{\Gamma}=N_{f}\dfrac{4g^{4}(M_{P}^{*})^{2}m_{\widetilde{G}}^{2}}{5\pi^{2}M_{W}^{4}}~\dfrac{\xi_{L}^{2}+\xi_{R}^{2}}{\kappa_{Z\widetilde{G}}}\left( \dfrac{|\Delta m_{+0}|}{m_{\text{NLSP}}} \right)^{5}~,
\end{equation}
where $N_{f}=4$ is the number of fermionic degrees of freedom the $\widetilde{\rho}_{1}^{\pm}$ decays to (those of $\widetilde{G}$). In the higgsino limit of the MRSSM and with $\widetilde{\chi}_{1}^{0}$ as the NNLSP, $(\xi_{L}^{2}+\xi_{R}^{2})=1/2$.
\\

\section{One-loop Higgs mass}
\label{app:Hmixing}

This appendix details the tree-level Higgs mass mixing entries and the one-loop $m_{h}$, as previously described in Refs.\cite{Bertuzzo:2014bwa,Diessner:2014ksa}. Neglecting terms containing the vevs of the adjoints (due to the hierarchies $M^{D},\mu_{u,d}\gg v_{\text{adj}}$ set by EWPT) the entries of $\mathcal{M}_{CP\text{-even}}^{2}$ are, in the basis  $\{ H_{d}^{0},H_{u}^{0},\phi_{\widetilde{B}},\phi_{\widetilde{W}}^{0}\}$,
\begin{align}
(\mathcal{M}_{\text{CP-even}}^{2})_{2,3} &\approx v~s_{\beta}\left[ g_{1}M_{\widetilde{B}}^{D}+\sqrt{2}\lambda_{\widetilde{B}}^{u}\mu_{u} \right] \notag \\
(\mathcal{M}_{\text{CP-even}}^{2})_{2,4} &\approx v~s_{\beta}\left[ -g_{2}M_{\widetilde{W}}^{D}-\lambda_{\widetilde{W}}^{u}\mu_{u} \right]~, \label{eq:mixingHuAdj}
\end{align}
which encodes the mixing of $H_{u}^{0}$, and
\begin{align}
(\mathcal{M}_{\text{CP-even}}^{2})_{1,3} &\approx v~c_{\beta}\left[ -g_{1}M_{\widetilde{B}}^{D}+\sqrt{2}\lambda_{\widetilde{B}}^{d}\mu_{d} \right] \notag \\
(\mathcal{M}_{\text{CP-even}}^{2})_{1,4} &\approx v~c_{\beta}\left[ g_{2}M_{\widetilde{W}}^{D}-\lambda_{\widetilde{W}}^{d}\mu_{d} \right]~, \label{eq:mixingHdAdj}
\end{align}
that describes the mixing of $H_{d}^{0}$. The $\phi_{a}$ stand respectively for the real components of the adjoint $A_{a}$ scalar. As pointed out in \cite{Bertuzzo:2014bwa} an immediate way to increase the lightest eigenvalue consists of choosing the $\lambda_{\widetilde{B},\widetilde{W}}$ such that both terms inside the square brackets in each entry above carry opposite signs and cancel each other. Then Eqs. \ref{eq:mixingHuAdj} and \ref{eq:mixingHdAdj} require that
\begin{equation*}
\lambda_{\widetilde{B}}^{u},\lambda_{\widetilde{W}}^{u}<0\text{~~~~~and~~~~~}\lambda_{\widetilde{B}}^{d},\lambda_{\widetilde{W}}^{d}>0
\end{equation*}
to favor the mentioned cancellation in all four entries.

The tree-level quartic of the Higgs receives one-loop corrections that are controlled by the adjoint scalar masses, the Dirac gaugino masses and the superpotential couplings $\lambda_{a}^{u,d}$. Including the non-negative, tractable expression for the loop piece of the Higgs quartic obtained via effective potential techniques for $(M^{D})^{2}\leq m_{\text{adj}}^{2}$ in \cite{Bertuzzo:2014bwa}, the SM-like Higgs mass is approximated in the large $\tan \beta$ limit by
\begin{align}
&m_{h}^{2} 
 \approx M_{Z}^{2} \notag \\
&- v^{2}\left[ \dfrac{(   g_{1}M_{\widetilde{B}}^{D}+\sqrt{2}\lambda_{\widetilde{B}}^{u} \mu_{u}  )^{2}}{4(M_{\widetilde{B}}^{D})^{2}+m_{\text{adj}}^{2}+2B_{\widetilde{B}}}+\dfrac{(   g_{2}M_{\widetilde{W}}^{D}+\lambda_{\widetilde{W}}^{u} \mu_{u}   )^{2}}{4(M_{\widetilde{W}}^{D})^{2}+m_{\text{adj}}^{2}+2B_{\widetilde{W}}} \right] \notag \\
&+ 2v^{2}\left( \dfrac{5(\lambda_{\widetilde{W}}/\sqrt{2})^{2}+2(\lambda_{\widetilde{W}}/\sqrt{2})^{2}\lambda_{\widetilde{B}}^{2}+\lambda_{\widetilde{B}}^{2}}{16\pi^{2}} \right.  \notag \\
&\times \log{\left[ \dfrac{m_{\text{adj}}^{2}}{(M^{D})^{2}} \right]}+ \left. \dfrac{(\lambda_{\widetilde{W}}/\sqrt{2})^{2}\lambda_{\widetilde{B}}^{2}}{16\pi^{2}} \right)+\bigl. \delta m_{h}^{2}\bigr|_{\text{stops}} \label{eq:CWpotential}
\end{align}
 where $M^{D}$ and $m_{\text{adj}}^{2}$ are common gaugino and adjoint masses. Regardless of the combined $D$- and $F$-breaking origin of its soft mass,  the stop one-loop correction to $m_{h}^{2}$ is parametrically the same as for the MSSM,
\begin{equation}
\bigl. \delta m_{h}^{2}\bigr|_{\text{stops}}=3\cdot \frac{y_{t}^{2}m_{t}^{2}}{4\pi^{2}}\log{\dfrac{m_{\widetilde{t}}^{2}}{m_{t}^{2}}}~. \label{eq:stopLoop}
\end{equation}

\section{Fine-tuning contributions}
\label{app:FT}

The leading contributions to $\Delta_{v}$ (the vev fine-tuning) are listed below and are quoted from \cite{Bertuzzo:2014bwa} and \cite{Gherghetta:2012gb}
\begin{align}
                   \Delta_{\mu} &\simeq \dfrac{4\mu^{2}}{m_{Z}^{2}}~, \\
\Delta_{\widetilde{W}} &\simeq \left| \dfrac{3\bigl( \lambda_{\widetilde{W}}/\sqrt{2} \bigr)^{2}m_{\text{adj}}^{2}}{4\pi^{2}m_{Z}^{2}}~L\right|~, \\
\Delta_{\widetilde{B}} &\simeq \left| \dfrac{\lambda_{\widetilde{B}}^{2}m_{\text{adj}}^{2}}{4\pi^{2}m_{Z}^{2}}~L\right|~, \\
\Delta_{R} &\simeq \left| \dfrac{\bigl[ \lambda_{\widetilde{B}}^{2}+3\bigl( \lambda_{\widetilde{W}}/\sqrt{2} \bigr)^{2} \bigr]m_{R}^{2}}{16\pi^{2}m_{Z}^{2}}~L\right|~, \\
\Delta_{\widetilde{t}} &\simeq \left| \dfrac{3y_{t}^{2}m_{\widetilde{t}}^{2}}{4\pi^{2}m_{Z}^{2}}~L\right|~,
\end{align}
where $L\equiv \log{(\Lambda_{\text{mess}}/m_{\widetilde{t}})}$. The $\Delta_{i}$'s above receive small $O(v^{2}/m_{\text{new}}^{2})$ corrections where $m_{\text{new}}$ stands for a combination of Dirac gaugino, soft adjoint masses, and $B_{\text{adj}}$-terms.

\bibliographystyle{apsrev4-1}
\bibliography{allReferences.bib}

\end{document}